\documentclass[cha,aip,reprint,noshowpacs,superscriptaddress,floatfix,letterpaper,longbibliography]{revtex4-2}
\usepackage{amsmath,amssymb,amsfonts,mathtools,bbm,bm,bbold}
\usepackage{newtxtext}
\usepackage{physics}
\usepackage[dvipsnames]{xcolor}
\definecolor{LapisLazuli}{RGB}{47, 102, 169}
\definecolor{my_blue}{HTML}{0072BD} 
\definecolor{my_red}{HTML}{cf1f19} 
\usepackage[hypertexnames=false,colorlinks=true,citecolor=LapisLazuli,linkcolor=LapisLazuli,urlcolor=LapisLazuli]{hyperref}
\usepackage{pgfplots}
\usetikzlibrary{arrows,intersections}
\pgfplotsset{compat = newest}
\usepackage{tikz-3dplot}
\usepackage{tikz}
\newcommand{\bX}{{\boldsymbol{X}}}
\newcommand{\bY}{{\boldsymbol{Y}}}
\newcommand{\stability}{{\bm{A}}}
\newcommand{\jacobian}{{\bm{M}}}
\newcommand{\ex}{{\bm{x}}}
\newcommand{\bphi}{{\bm{\phi}}}

\newcommand{\yu}{{\bm{u}}}
\newcommand{\brho}{{\bm{\varrho}}}
\newcommand{\bxi}{{\bm{\xi}}}
\newcommand{\stabilityH}{{\bm{\mathcal{H}}}}
\newcommand{\Hamiltonian}{H}
\newcommand{\bmP}{{\bm{p}}}
\newcommand{\bmQ}{{\bm{q}}}
\graphicspath{{../pure/PRX_sub/}}
\usepackage{multirow}

\begin{document}

\title{Phase space volume preserving dynamics for deterministic dynamical systems}

\author{Swetamber~Das}
\affiliation{Department of Physics, 
    SRM University -- AP, 
    Amaravati, Andhra Pradesh 522240, India
}
\author{Jason~R.~Green}
\email[]{jason.green@umb.edu}
\affiliation{Department of Chemistry,\
	University of Massachusetts Boston,\
	Boston, MA 02125, USA
}
\affiliation{Department of Physics,\
	University of Massachusetts Boston,\
	Boston, MA 02125, USA
}

\date{\today}
\begin{abstract}

Infinitesimal volumes stretch and contract as they coevolve with classical phase space trajectories according to linearized dynamics.
Unless these tangent-space dynamics are modified, chaotic evolution causes the
volume spanned by evolving tangent vectors to collapse. However, this collapse is unphysical and due to their exponential alignment along the most expanding direction, independent of the compressibility of the phase-space volume.
Here, we propose an alternative linearized dynamics and rectify the generalized Liouville equation to preserve phase space volume, even for non-Hamiltonian systems.
Within a classical density matrix theory, we define the time evolution operator from the anti-symmetric part of the stability matrix so that phase space volume is time-invariant.
The operator generates orthogonal transformations without distorting volume elements, providing an invariant measure for dissipative dynamics and a evolution equation for the density matrix akin to the quantum mechanical Liouville-von Neumann equation.
The compressibility of volume elements is determined by a non-orthogonal operator made from the symmetric part of the stability matrix.
We analyze complete sets of basis vectors for the tangent space dynamics of chaotic systems, which may be dissipative, transient or driven, without re-orthogonalization of tangent vectors.
The linear harmonic oscillator, the Lorenz-Fetter model, and the H\'enon-Heiles system demonstrate the computation of the instantaneous Lyapunov exponent spectrum and the local Gibbs entropy flow rate using these bases and show that it is numerically convenient.

\end{abstract}

\maketitle

\begin{quotation}
In chaotic systems, from planetary motion to chemical reactions, the phase space volume spanned by an initially localized set of tangent vectors collapses over time due tocontinuous stretching and folding--an artifact that is a challenge in the analysis of phase-space dynamics.
Hamiltonian systems naturally conserve volume, and many physical systems are driven or dissipative, for which the volume may expand or contract.
 We introduce a general framework that redefines the underlying linearized evolution and prevents the unphysical vanishing of phase-space volume in chaotic dynamics when tangent vectors align with the most expanding direction.
By constructing a time-evolution operator from the antisymmetric part of the stability matrix, we establish a volume-preserving evolution in tangent space governed by a classical analogue of the Liouville-von Neumann equation.
These operators provide an invariant measure that conserves phase-space volume and can generate a continuously orthogonal tangent frame, enabling a systematic evaluation of local Lyapunov exponents in both conservative and dissipative systems.
The theory offer a geometric perspective of the distinct roles of local stretching and rotation in Lyapunov stability analysis.
\end{quotation}

\section{Introduction}

Phase space volume conservation is a defining feature of Hamiltonian dynamics~\cite{Gibbs_2010}.
This time invariance of phase space volume is central to the Liouville's theorem for divergence-free phase space flows.
A more rigorous statement considers the symplectic geometry of the Hamiltonian flow  which leads to the incompressiblity of phase space volumes~\cite{arnold1989mathematical}.
Incompressibility is usually understood as the preservation of an infinitesimal volume element evolving in time despite distortion (expansion, contraction, and rotation) by the governing flow~\cite{tolman1979principles}.
However, many dynamical systems of interest are characterized by phase space \emph{variability}.
Systems that are dissipative, driven, or constrained do not conserve phase space volume.
The statistical mechanics of these non-Hamiltonian systems must then be based on generalizations of Liouville's theorem and Liouville's equation.
The most straightforward generalization~\cite{Steeb1979,Andrey1985a,Andrey1985b} of Liouville's equation for an arbitrary deterministic system invokes the continuity equation $\partial_t \rho + \boldsymbol{\nabla}\cdot(\rho \boldsymbol f) = 0$ for the phase space density $\rho(\ex,t)$. 
This density corresponds to an ensemble of systems evolving according to a dynamical flow $\dot{\ex}=\boldsymbol f(\ex)$. 
Because Liouville's equation also holds for out of equilibrium systems, its solution i.e., the phase space probability density $\rho(\ex,t)$ is used to compute mean phase space observables such as Gibbs entropy rate.

For dissipative systems,  the rate at which the phase space probability density $\rho(t)$ varies with time is determined by the divergence of the flow $\boldsymbol{\nabla}\cdot \dot \ex$.
This divergence also measures the compressibility of the phase space, which is the contraction rate of phase space volume element $\delta V(t)$ for dissipative dynamical systems. 
For non-equilibrium steady states, $\rho$ diverges as $\delta V$ collapses onto a attractor with fractal dimension~\cite{Hoover1995,Klages2007}. 
As a consequence, irreversible-thermodynamics concepts such as nonequilibrium Gibbs entropy rate becomes undefined. 
The non-smoothness of the $\rho$ in non-Hamiltonian systems due to phase space variability has been a long-standing concern~\cite{Ramshaw1986, Ramshaw_2002,Tuckerman_1999, TucLiuGioMar2001,  Ezra2004}. 
This ambiguity may be resolved either by course-graining phase space or by determining a suitable invariant measure~\cite{Ramshaw1986}. 
The presence of an appropriate invariant measure implies that the phase space average of a specific property can be correlated with the time-average of that same property along a trajectory (assuming ergodicity). To derive an appropriate measure, a typical way is to modify the phase space probability density by including either a geometric factor or a ``measure function" to normalize the probability density $\rho(\ex,t)$~\cite{Tuckerman1997,Tuckerman_1999,TucLiuGioMar2001, Ezra2004,Ramshaw_2019, Ramshaw_2020, Ramshaw_2022}. 
Obtaining such smooth invariant measures that satisfy the Liouville equation is a mathematically nontrivial and often subtle for non-Hamiltonian systems~\cite{holian1989nonlinear, ruelle1999smooth,dorfman1999introduction}.

Despite significant theoretical progress, computing the phase space density for large systems remains inherently challenging due to dynamical chaos. When phase space trajectories diverge at an exponential rate, the phase space volume develops intricate fine structures, causing $\rho$ to become non-smooth and rendering Liouville's equation inapplicable. This challenge persists even in chaotic Hamiltonian systems, where the conservation of phase space volume does not guarantee a smooth probability density due to continuous stretching and folding of phase space volume. As a result, the exponential separation of nearby trajectories renders direct computation of a well-behaved phase-space density computationally intensive.
The exponential separation also causes tangent vectors to rapidly align with the most expanding direction, making phase-space volume computation difficult~\cite{Skokos2016}. 
Here and throughout, this ``collapse'' refers to the phase space volume represented by a finite set of evolving tangent vectors due to dynamical alignment, and not to a violation of phase-space volume conservation of the underlying flow.
The issue is conventionally addressed through Gram-Schmidt orthogonalization~\cite{Ott_2002}, but its iterative nature is computationally expensive and suffers from loss of orthogonality due to round-off errors~\cite{Leon2013,Ford2015}. 
This challenge is even more pronounced in non-ergodic chaotic systems, where complex phase-space structures and high-dimensional dynamics further complicate computations~\cite{Zachary2008, hahn2021}.

Here, we address two questions in the context of classical chaotic dynamics: (1) how to generalize Liouville's equation to obtain a well-defined invariant measure and (2) how local Lyapunov exponents, in various bases, contribute to the local Gibbs entropy flow rate.
To do that, we use the emerging framework of classical density matrix theory~\cite{dasDensityMatrixFormulation2022}, which is akin to quantum density matrix formulation in its mathematical treatment. 
In Sec.~\ref{sec:den-mat}, we summarize the theory and show how quantities like the instantaneous Lyapunov exponents and local phase space volume contraction rates appear.
In Sec.~\ref{sec:evolution}, we obtain a set of evolution operators for a modified tangent space dynamics.
Based on the mathematical properties of these operators we derive, in Sec.~\ref{sec:GLE/T}, a set of generalized Liouville's theorem and equations.
To describe local rates of stretching and contraction of phase space volume elements, we introduce several ``basis states" in Sec.~\ref{sec:basis} and show how to derive those local rates for Hamiltonian and non-Hamiltonian systems.
We compute these rates for a several model systems including the damped harmonic oscillator, Lorenz-Fetter model, and H\'enon-Heiles system in Sec.
\ref{sec:case_examples}.

\section{Density matrix theory}\label{sec:den-mat}

We recently introduced a classical density matrix theory for deterministic dynamical systems to obtain upper and lower bounds on finite-time Lyapunov exponents and a family of classical speed limits on its time evolution~\cite{dasDensityMatrixFormulation2022,dasSpeedLimitsDeterministic2023,DasMax2024,DasGreen-spectral-bounds}.
The theory is based on a projection operator $\brho$ defined as the outer product of unit tangent vectors, $\brho := \dyad{\delta \yu(t)}{\delta \yu(t)}$,
which acts as a classical density matrix. 
Its norm-preserving evolution is
\begin{align}\label{eq:EOM_rho}
	d_t\brho = \{\bar\stability_+,\brho\}  + [\stability_-,\brho].
\end{align}
Here, $\bar\stability_+ = \stability_+ - \langle\stability_+\rangle= \stability_+ - \Tr(\stability_+\brho)$ and $\stability_\pm$ represent the symmetric and anti-symmetric parts of the stability matrix $\stability$ appearing in the anti-commutator $\{\bX,\bY\} = \bX\bY + \bY\bX$ and the commutator $[\bX,\bY] = \bX\bY - \bY\bX$, respectively.
We refer to $\brho$ as a ``pure state'' because it is constructed from a single tangent vector $\ket{\delta\yu}$. 
The norm preserving property of the dynamics manifests as the time invariance of the trace: $\Tr \dot \brho = 0$.
Equation \ref{eq:EOM_rho} also shows that the dynamics of $\brho$ is a sum of two norm preserving dynamics
\begin{align}
\underbrace{\{\bar\stability_+,\brho\}}_\text{incoherent}, \quad \text{and} \quad  \underbrace{ [\stability_-,\brho]}_\text{unitary}.
\end{align}
The symmetric $\stability_+$ and antisymmetric $\stability_-$ parts of the stability matrix evolve tangent vectors differently: The antisymmetric part is a unitary (orthogonal) evolution operator and the incoherent part stretches/contracts tangent vectors.

To see that the incoherent part stretches and contracts tangent vectors, we identify the ``expectation value'' $\langle \stability_+ \rangle$ appearing in the anti-commutator in Eq.~\ref{eq:EOM_rho}, as the local Lyapunov exponent corresponding to the density matrix $\brho$:
\begin{align}
	r = \Tr(\stability_+\brho) = \bra{\delta \yu}\stability_+\ket{\delta \yu}.
\end{align} 
The time average of this local stretching rate of a tangent vector $\ket{\delta \yu}$ is
\begin{align}
	\lambda(t) := \lambda(t,t_0) = |t-t_0|^{-1}\int_{t_0}^tr(t)\,dt,
\end{align}
the finite-time Lyapunov exponent.
The usual phase-space invariant Lyapunov exponent is obtained by taking the asymptotic limit, $\lambda=\lim_{t\to\infty}\lambda(t)$.
    Finite-time exponents can be defined for a unit tangent vector $\ket{\delta \yu}$ in an arbitrary direction and are not restricted to the most expanding directions commonly used in standard Lyapunov analyses (e.g., the tangent-space direction associated with the largest instantaneous growth rate).
They remain well defined even in purely contracting dynamics, where all instantaneous growth rates are negative.
In chaotic systems, which are the primary focus here, the most expanding direction corresponds to a locally expanding vector.
One may initially choose a tangent vector aligned with this most expanding direction.
However, under $\jacobian_-$ dynamics this vector undergoes continuous rotation, and therefore need not remain aligned with the instantaneous most expanding direction at later times.
This contrasts with the standard Gram-Schmidt (or QR) procedure, which evolves the full linearized dynamics and continually adjusts the orthonormal frame so that one vector tracks the dominant stretching direction.

Together, the incoherent (symmetric) and unitary (anti-symmetric) contributions determine the tangent-space dynamics of a deterministic system, encoding
local stretching, contraction, and rotation of perturbations along a trajectory.
In the following section, we construct explicit evolution operators associated
with these two contributions and analyze their mathematical properties, which
will allow us to formulate generalized Liouville equation for both Hamiltonian and non-Hamiltonian systems.

\section{Evolution operators}\label{sec:evolution}
 
\begin{figure}[t!]
	\centering
	\begin{tikzpicture}[thick, >=stealth]
	\definecolor{S_Gold}{RGB}{255,192,0}
	\hspace*{0cm}
	\draw[Black,very thick,-latex] (-2.5, -0.9,  -2.0) to[bend left] (4.0,-3.8,-2.89);
	\draw [->,>=stealth,thick, line width=1pt] (0,0,0) -- (1,1.05,0.85) node [at end, right] {\small $\,\ket{\delta \yu_1(t_0)}$};
\draw [->,>=stealth,thick, line width=1pt]
  (0,0,0) -- (0.98,0.22975,0,0.5)
  node [at end, right] {\small $\,\ket{\delta \yu_2(t_0)}$};
\draw[black, thick, ->, >=stealth]
  (0.4,0.09,0) arc[start angle=30, end angle=66, radius=0.4]
  node[midway, right, xshift=1.1pt,yshift=3.5pt] {\small $\theta$};
	\tdplotsetmaincoords{180}{0}
	\tdplotsetrotatedcoords{0}{0}{0}
	\node[] (O) at (0,0,0){};
	\begin{scope}[rotate=0, xscale=1, yscale=1, shift={(0.0,0.0)}]
	\coordinate (O) at (0,0);
	\shade[ball color=lightgray!90!black,opacity=0.6] (0,0) coordinate(Hp) circle (1) ;
	\end{scope}
	\hspace*{3cm}
	\draw [->,>=stealth,thick, line width=1pt, color=my_red] (0,-0.8,0) -- (0.78,-1.0,-0.55) node [at end, right,yshift=4pt,xshift=-1.2] {\small $\,\jacobian_-\ket{\delta \yu_2(t_0)}$};
	\draw [->,>=stealth,thick, line width=1pt, color=my_red]
  (0,-0.8,0) -- (0.657,-0.49,-0.55)
  node [at end, right,xshift=2.5pt,yshift=2pt] {\small $\,\jacobian_-\,\ket{\delta \yu_1(t_0)}$};
\draw[black, thick, ->, >=stealth]
  (0.4,-0.81,0) arc[start angle=30, end angle=66, radius=0.4]
  node[midway, right, xshift=2pt,yshift=2.2pt] {\small $\theta$};

	\draw [->,>=stealth,thick, line width=1pt, color=my_blue] (0,-0.8,0) -- (0.7,-1.40,-0.55) node [at end, right, yshift=3.5pt] {\small $\,\widetilde\jacobian\ket{\delta \yu_1(t_0)}$};
	\draw [->,>=stealth,thick, line width=1pt, color=my_blue] (0,-0.8,0) -- (0.66,-1.50,-0.55) node [at end, right, yshift=-4.20pt] {\small $\,\widetilde\jacobian\ket{\delta \yu_2(t_0)}$};
	
	\tdplotsetmaincoords{180}{0}
	\tdplotsetrotatedcoords{0}{0}{0}
	\node[] (O) at (0,0,0){};
	\begin{scope}[rotate=0, xscale=1, yscale=1, shift={(0.0,-0.80)}]
	\coordinate (O) at (0,0);
	\shade[ball color=lightgray!90!black,opacity=0.6] (0,0) coordinate(Hp) circle (1) ;
	\end{scope}
	\end{tikzpicture}
\caption{\label{fig:evolve_per_ket}
Time evolution of two tangent vectors $\ket{\delta \yu_1(t_0)}$ and $\ket{\delta \yu_2(t_0)}$ at time $t_0$ to the states $\ket{\delta \yu_1(t)}$ and $\ket{\delta \yu_2(t)}$ at time $t$, under two norm-preserving dynamics governed by the evolution operators $\widetilde{\jacobian}$ (bottom) and $\jacobian_-$ (top) along a given classical trajectory.
The operator $\jacobian_-$ is orthogonal and therefore preserves both the norm and the angle $\theta$ between the two states (shown in red) over time.
While $\widetilde{\jacobian}$ preserves the norm, it does not preserve the angle: the two tangent vectors evolve to become increasingly aligned, causing the angle between them to decrease as they approach the most expanding direction along the trajectory.
}

\end{figure}

The solution of Eq.~\ref{eq:EOM_rho}
\begin{align}
	\brho(t) = \widetilde{\jacobian}\brho(t_0)\widetilde{\jacobian}^\top
\end{align}
evolves the density matrix from time $t_0$ to time $t$ with the operator
\begin{align}\label{eq:expression_M}
\widetilde{\jacobian}(t,t_0) =\frac{ \mathcal{T}e^{\int_{t_0}^t\,\stability_-dt'}\mathcal{T}e^{\int_{t_0}^t\,\stability_+dt'}}{\langle\mathcal{T}e^{2\int_{t_0}^t\stability_+dt'}\rangle^{1/2}}
\end{align}
where $\mathcal{T}$ is the time ordering operator.
Figure~\ref{fig:evolve_per_ket} shows a schematic illustrating the action of $\widetilde{\jacobian}$ on the tangent vector $\ket{\delta \yu}$.
This evolution operator $\widetilde{\jacobian}$ is reminiscent of the Dyson series that appears in the interaction picture in quantum mechanics~\cite{Joachain1983}.
It has several properties, proven in Appendix~\ref{SI:evol-mat}:
\begin{enumerate}
	\item Norm-preserving: $ \bra{\delta \yu(t_0)}\widetilde{\jacobian}^\top\widetilde{\jacobian}\ket{\delta \yu(t_0)} = \bra{\delta \yu(t)}\ket{\delta \yu(t)} =1$ 
		\item Normal : $\widetilde{\jacobian}\widetilde{\jacobian}^\top = \widetilde{\jacobian}^\top\widetilde{\jacobian}$
	\item Non-unitary:  $\widetilde{\jacobian}^\top\widetilde{\jacobian} \neq \mathbb{1}$
\end{enumerate}
It can be decomposed into two normal operators:
\begin{align}
	\jacobian_- = \mathcal{T}e^{\int_{t_0}^t\,\stability_-dt'}, \quad \widetilde{\jacobian}_+ = \frac{\mathcal{T}e^{\int_{t_0}^t\,\stability_+dt'}}{\langle\mathcal{T}e^{2\int_{t_0}^t\stability_+dt'}\rangle^{1/2}}, 
\end{align}
so that $\jacobian_-\widetilde{\jacobian}_+ = \widetilde{\jacobian}$.
Any stretching of a $\ket{\delta \yu}$ is due to $\widetilde{\jacobian}_+$, which involves the symmetric part of the stability matrix $\stability_+$,
and evolves a pure state as
\begin{align}
\brho_{+}(t) = \widetilde{\jacobian}_+\brho(t_0)\widetilde{\jacobian}_+.
\end{align}
For example, a natural choice of basis vectors for expressing $\brho$ is the eigenvectors of $\stability_+$ (Sec.~\ref{sec:basis}).

While $\jacobian_-$ is independent of any tangent vector, both operators $\widetilde{\jacobian}$ and $\widetilde{\jacobian_+}$ are state dependent.
Their denominators include an expectation value over a given the tangent vector $\ket{\delta \yu}$.
The tilde in $\widetilde{\jacobian}$ and $\widetilde{\jacobian}_+$ indicates their dependence on the tangent vector.
Both these operators are specific to pure states and preserve the norm of the state.

Because the operators $\widetilde{\jacobian}$ and $\widetilde{\jacobian_+}$ define norm-preserving dynamics of tangent vectors, they conserve phase space volumes over time, even if the dynamics are non-Hamiltonian.
However, these evolution operators are not unitary (orthogonal) and do not preserve the initial angle between any two tangent vectors.
So, without repeated reorthogonalization, the tangent vectors eventually collapse to the most expanding direction along any chaotic trajectory causing the phase space volume spanned by these vectors to collapse for both Hamiltonian and non-Hamiltonian systems~\cite{Wolf1985}.
As we show next, the operator $\jacobian_-$ prevents this collapse.

\subsection*{Unitary evolution of a pure state}
\begin{figure}[!t!]
	\centering
	\textbf{Tangent space evolution: $\widetilde{\jacobian}$ dynamics}\\
	\vspace{0.4cm}
	\begin{tikzpicture}[thick, >=stealth]
	\definecolor{S_Gold}{RGB}{255,192,0}
	\vspace*{1cm}
	\draw[Black,very thick,-latex] (-2.5, -0.9,  -2.0) to[bend left] (4.0,-3.85,-2.89);
	\draw [->,>=stealth,thick, line width=1pt] (0,-0.03,0) -- (1,1.05,0.85) node [at end, right] {\small $\,\ket{\delta \yu_j(t_0)}$};
	\draw [->,>=stealth,thick, line width=1pt] (0,-0.03,0) -- (-0.32,1.1,1.25) node [at end, left] {\small $\,\ket{\delta \yu_i(t_0)}$};
	\draw [->,>=stealth,thick, line width=1pt] (0,-0.03,0) -- (-0.9,-1.85,-2.25) node [at end, below] {\small $\,\ket{\delta \yu_k(t_0)}$};;
	\tdplotsetmaincoords{180}{0}
	\tdplotsetrotatedcoords{0}{0}{0}
	\node[] (O) at (0,0,0){};
	\begin{scope}[rotate=0, xscale=1, yscale=1, shift={(0.0,0.0)}]
	\coordinate (O) at (0,0);
	\shade[ball color=lightgray!90!black,opacity=0.6] (0,0) coordinate(Hp) circle (1) ;
	\end{scope}
	\hspace*{3cm}
	\draw [->,>=stealth,thick, line width=1pt, color=my_blue] (-0.1,-0.8,0) -- (0.8,-1.0,-0.50) node [at end, right] {\small $\,\ket{\delta \yu_i(t)}$};
	\draw [->,>=stealth,thick, line width=1pt, color=my_blue] (-0.1,-0.8,0) -- (0.7,-1.40,-0.55) node [at end, right] {\small $\,\ket{\delta \yu_j(t)}$};;
	\draw [->,>=stealth,thick, line width=1pt, color=my_blue] (-0.1,-0.8,0) -- (0.76,-1.42,.50) node [at end, below] {\small $\quad\ket{\delta \yu_k(t)}$};;
	\tdplotsetmaincoords{180}{0}
	\tdplotsetrotatedcoords{0}{0}{0}
	\node[] (O) at (0,0,0){};
	\begin{scope}[rotate=0, xscale=1, yscale=1, shift={(0.0,-0.80)}]
	\coordinate (O) at (0,0);
	\shade[ball color=lightgray!90!black,opacity=0.6] (0,0) coordinate(Hp) circle (1) ;
	\end{scope}
	\end{tikzpicture}
	
	\centering
	\textbf{Unitary evolution: $\jacobian_-$ dynamics}\\
	\vspace{0.4cm}
	\begin{tikzpicture}[thick, >=stealth]
	\definecolor{S_Gold}{RGB}{255,192,0}
	\hspace*{0cm}
	\draw[Black,very thick,-latex] (-2.5, -0.9,  -2.0) to[bend left] (4.0,-3.85,-2.89);
	\draw [->,>=stealth,thick, line width=1pt] (0,-0.03,0) -- (1,1.05,0.85) node [at end, right] {\small $\,\ket{\delta \yu_j(t_0)}$};
	\draw [->,>=stealth,thick, line width=1pt] (0,-0.03,0) -- (-0.32,1.1,1.25) node [at end, left] {\small $\,\ket{\delta \yu_i(t_0)}$};
	\draw [->,>=stealth,thick, line width=1pt] (0,-0.03,0) -- (-0.9,-1.85,-2.25) node [at end, below] {\small $\,\ket{\delta \yu_k(t_0)}$};
	\tdplotsetmaincoords{180}{0}
	\tdplotsetrotatedcoords{0}{0}{0}
	\node[] (O) at (0,0,0){};
	\begin{scope}[rotate=0, xscale=1, yscale=1, shift={(0.0,0.0)}]
	\coordinate (O) at (0,0);
	\shade[ball color=lightgray!90!black,opacity=0.6] (0,0) coordinate(Hp) circle (1) ;
	\end{scope}
	\hspace*{3cm}
	\draw [->,>=stealth,thick, line width=1pt, color=my_red] (-0.1,-0.8,0) -- (0.1,0.27,0.20) node [at end, above] {\small $\,\ket{\delta \yu_i(t)}$};
	\draw [->,>=stealth,thick, line width=1pt, color=my_red] (-0.1,-0.8,0) -- (0.7,-1.40,-0.55) node [at end, right] {\small $\,\ket{\delta \yu_j(t)}$};
	\draw [->,>=stealth,thick, line width=1pt, color=my_red] (-0.1,-0.8,0) -- (-0.55,-1.3,.50) node [at end, left] {\small $\,\ket{\delta \yu_k(t)}$};
	\tdplotsetmaincoords{180}{0}
	\tdplotsetrotatedcoords{0}{0}{0}
	\node[] (O) at (0,0,0){};
	\begin{scope}[rotate=0, xscale=1, yscale=1, shift={(0.0,-0.80)}]
	\coordinate (O) at (0,0);
	\shade[ball color=lightgray!90!black,opacity=0.6] (0,0) coordinate(Hp) circle (1) ;
	\end{scope}
	\end{tikzpicture}
	\caption{\label{fig:evolve_basis_ket} Time evolution of a set of basis states from an initial time $t_0$ to a time $t$ under two norm-preserving dynamics, $\widetilde{\jacobian}$ and $\jacobian_-$ along a given classical trajectory. However, unlike $\widetilde{\jacobian}$, the operator $\jacobian_-$ preserves the angles between the basis states. Their inner products are time invariant: $\bra{\delta \yu_i(t)}\ket{\delta \yu_j(t)} = \bra{\delta \yu_i(t_0)}\ket{\delta \yu_j(t_0)}$ under unitary evolution $\jacobian_-$.}
\end{figure}

Unlike $\widetilde{\jacobian}$ and $\widetilde{\jacobian_+}$, the operator $\jacobian_-$ generates a unitary evolution --- it preserves the inner product of two pure states:
\begin{align}
\bra{\delta \yu_i(t)}\ket{\delta \yu_j(t)} &= \bra{\delta \yu_i(t_0)}\jacobian_-^\top\jacobian_-\ket{{\delta\yu_j(t_0)}}\\
&= \bra{\delta \yu_i(t_0)}\ket{\delta \yu_j(t_0)},
\end{align}
as $\jacobian_-^\top\jacobian_- = \mathbb{1}$.
The operator belongs to $SU(n)$ as its determinant is one.
Figure~\ref{fig:evolve_basis_ket} shows the action of $\jacobian_-$ on a set of three mutually perpendicular pure states along a chaotic trajectory of the Lorenz attractor.

In dynamical systems with a time-independent $\stability$, for instance, the linear harmonic oscillator or Chua's circuit~\cite{matsumoto1985}, the operator becomes $\jacobian_- = e^{\stability_-t}$.
It is a classical analogue of the evolution operator $\hat{\boldsymbol{U}} = e^{-i\hat{\boldsymbol{H}}t}$ in quantum mechanics producing unitary evolution of a quantum state for a time-independent Hamiltonian $\hat{\boldsymbol{H}}$. With this analogy, the equation of motion of a pure state evolving under $\jacobian_-$ is
\begin{align}
	d_t\brho_- = [\stability_-,\brho_-],
\end{align}
which has the solution
\begin{align}
	\brho_-(t) = \jacobian_-^\top\brho(t_0)\jacobian_-.
\end{align}
For $\jacobian_-$ dynamics, a natural basis to write $\brho_-$ is the set of complex eigenvectors of $\stability_-$. We discuss this case of a complex tangent space in Sec.~\ref{sec:basis}.

The preserved inner product under $\jacobian_-$ implies that volumes are invariant under time evolution; that is, volumes defined by $k$ linearly independent vectors in a subspace of an $n$-dimensional tangent space, where $k\leq n$.
We use this feature of $\jacobian_-$ to arrive at a generalization of Liouville's theorem that avoids collapse of the tangent vectors in chaotic systems.

\section{A generalized Liouville's theorem and equation}\label{sec:GLE/T}

The determinant $|\brho|$ represents the local phase-space volume associated with a complete set of tangent-space basis vectors, and its equation of motion provides a generalized Liouville description. To see this geometrically, we consider a set of basis vectors $\{\ket{\delta\bm{\phi}_i(t)}\}_{i=1}^n$  spanning an $n$-dimensional phase space volume element $d\mathcal{V}(t)$ at time $t$.  The outer products of these basis vectors form a set of pure states  $\brho_i = \dyad{\delta\bm{\phi}_i}$. The normalized sum of these pure basis states with equal weights defines a ``maximally mixed'' state:
\begin{align}\label{eq:maximally_mixed}
\brho_\text{max} = \frac{1}{n} \sum_{i=1}^n \brho_i = \frac{1}{n} \boldsymbol V\boldsymbol V^\top.
\end{align}
Here $\boldsymbol{V}$ contains a set of complete basis vectors at time $t$ stacked as columns: $\boldsymbol{V} = [\ket{\delta \bm{\phi}_1} \, \ket{\delta \bm{\phi}_2} \, \cdots \quad\ket{\delta \bm{\phi}_n}]$. The volume spanned by the basis vectors is  $d\mathcal{V} = |\boldsymbol V|,$  and the determinant of $|\brho_\text{max}|$ satisfies: $|\brho_\text{max}| = n^{-n} |\boldsymbol V\boldsymbol V^\top| = n^{-n} |\boldsymbol V|^2.$  Rewriting, we obtain the geometric relation  $(d\mathcal{V})^2 = n^n |\brho_\text{max}|,$  which connects the phase space volume element $d\mathcal{V}$ to determinant  $|\brho_\text{max}|$.  This state can also be used to compute the \emph{local} phase space volume dissipation rate:
\begin{align}\label{eq:dissipation_rate}
	\Lambda(t) = n\Tr(\stability_+\brho_\text{max}) =\sum_{j=1}^n\Tr(\stability_+\brho_j).
\end{align} 
For Hamiltonian dynamics, $\Lambda(t) = 0$ and for dissipative systems, $\Lambda(t) < 0$. A set of basis states, therefore, provides the full spectrum of instantaneous or local Lyapunov exponents at a given time $t$ for an arbitrary dynamical system (Sec.~\ref{sec:basis}).

To track how the phase space volume $d\mathcal{V}$ changes over time, we express the dynamics of $\brho_\text{max}$ using the operator $\widetilde{\jacobian}'$ as
\begin{align}
	\brho_\text{max}(t) = \widetilde{\jacobian}'\brho_\text{max}(t_0)\widetilde{\jacobian}'^\top.
\end{align}
This operator $\widetilde{\jacobian}'$ is constructed as $\widetilde{\jacobian}' = \sum_{i,j=1}^n[\delta_{ij}]\widetilde{\jacobian}_i$. In Appendix~\ref{SI_vol}, we show that the determinant $|\widetilde{\jacobian}'| = 1$, indicating that it is a volume-preserving operator that belongs to $SL(n)$, and is generally non-orthogonal. The determinant of $\brho_\text{max}$ remains unchanged over time:
\begin{align}
|\brho_\text{max}(t)| = |\brho_\text{max}(t_0)|.
\end{align}
Thus, the operator $\widetilde{\jacobian}'$ generates a coordinate transformation that preserves the phase volume element between time $t_0$ and $t$:
\begin{align}
	d\mathcal{V}(t) = d\mathcal{V}(t_0).
\end{align}
This volume preservation can also be confirmed directly from the equation of motion Eq.~\ref{eq:EOM_rho} for the maximally mixed density matrix $\brho_{\mathrm{max}}$ (see Appendix~\ref{SI:LE-derivation}):
	\begin{align}\label{Eq:GLE}
	d_t\ln|\brho_{\mathrm{max}}|
	&= \Tr(d_t\ln \brho_{\mathrm{max}})
	= 0,
	\end{align}
provided $\brho_{\mathrm{max}}$ is invertible. The invertibility requirement excludes pure states whose determinant vanishes. The evolution law of the determinant $|\brho_{\mathrm{max}}|$ therefore holds for the maximally mixed density matrix constructed from a complete set of basis vectors, which represents an ensemble of phase-space points in the neighborhood of a reference trajectory (for details, see Ref.~\onlinecite{Sahbani2025}).
For a general mixed state $\brho$, its non-vanishing determinant $|\brho|$ defines a phase-space volume spanned by its column vectors, which form a complete set of linearly independent basis vectors.
In the present construction, determinant preservation under the generalized Liouville equation (Eq.~\ref{Eq:GLE}) applies specifically to the maximally mixed state $\brho_{\mathrm{max}}$ constructed from a complete basis, which represents a local phase-space volume element.
The invariance of $|\brho_{\mathrm{max}}|$ implies that this complete set of linearly independent vectors remains so during time evolution.

However, the generalization of Liouville's equation under $\widetilde{\jacobian}$ reveals a challenge for chaotic trajectories. Although $\widetilde{\jacobian}$ preserves phase-space volume, it is generally non-orthogonal, so it does not preserve angles between tangent vectors. As a result, in chaotic systems tangent vectors align exponentially fast with the most expanding direction, causing the volume represented by these vectors to collapse during explicit evolution under $\widetilde{\jacobian}$ dynamics~\cite{benettin1980lyapunov2}.
This collapse of the tangent vectors on the most expanding direction is a known technical bottleneck in computing the full Lyapunov spectrum for chaotic systems, where numerical methods impose orthogonality among tangent vectors.
This issue is typically addressed by periodic orthonormalization of the tangent vectors using the Gram-Schmidt procedure.
However, the method is prone to numerical instability due to the accumulation of round-off errors during repeated reorthogonalization, which can artificially inflate or suppress estimated growth rates over long trajectories~\cite{Geist1990}.
Furthermore, loss of orthogonality in finite-precision arithmetic results in unreliable perturbation vector norms, leading to spurious positive Lyapunov exponents~\cite{gencay1996identification} or convergence failures~\cite{Hassanaly2019}.

The dynamics generated by $\jacobian_-$ dynamics is free from the orthonormalization issue because it generates a unitary (orthogonal) evolution. An initial complete set of orthonormal basis states remain so for all times (see Fig.~\ref{fig:evolve_basis_ket}) even on a chaotic trajectory. Under $\jacobian_-$ dynamics, a density matrix (pure or mixed) evolves as
\begin{align}
d_t\brho = [\stability_-,\brho]\quad\quad (\jacobian_- \, \text{dynamics}). 
\end{align}
This is classical analogue of the von Neumann equation for isolated quantum systems.  Like the quantum evolution operator $\hat U$, the unitary (orthogonal) operator $\jacobian_-$  preserves  the angle between any two pure states (tangent vectors) for all times. Consequently, the evolving perturbation vectors no longer collapse onto the most expanding direction. Also, because trace $\Tr \stability_- = 0$, 
	the corresponding Liouville's equation for the maximally mixed state remains unchanged: $d_t \ln |\brho_{\mathrm{max}}| = 0$. The phase-space volume represented by the determinant $|\brho_{\mathrm{max}}|$ is time invariant.

\begin{figure}[!t]
	\hspace{-0.5cm}\includegraphics[width=0.52\textwidth]{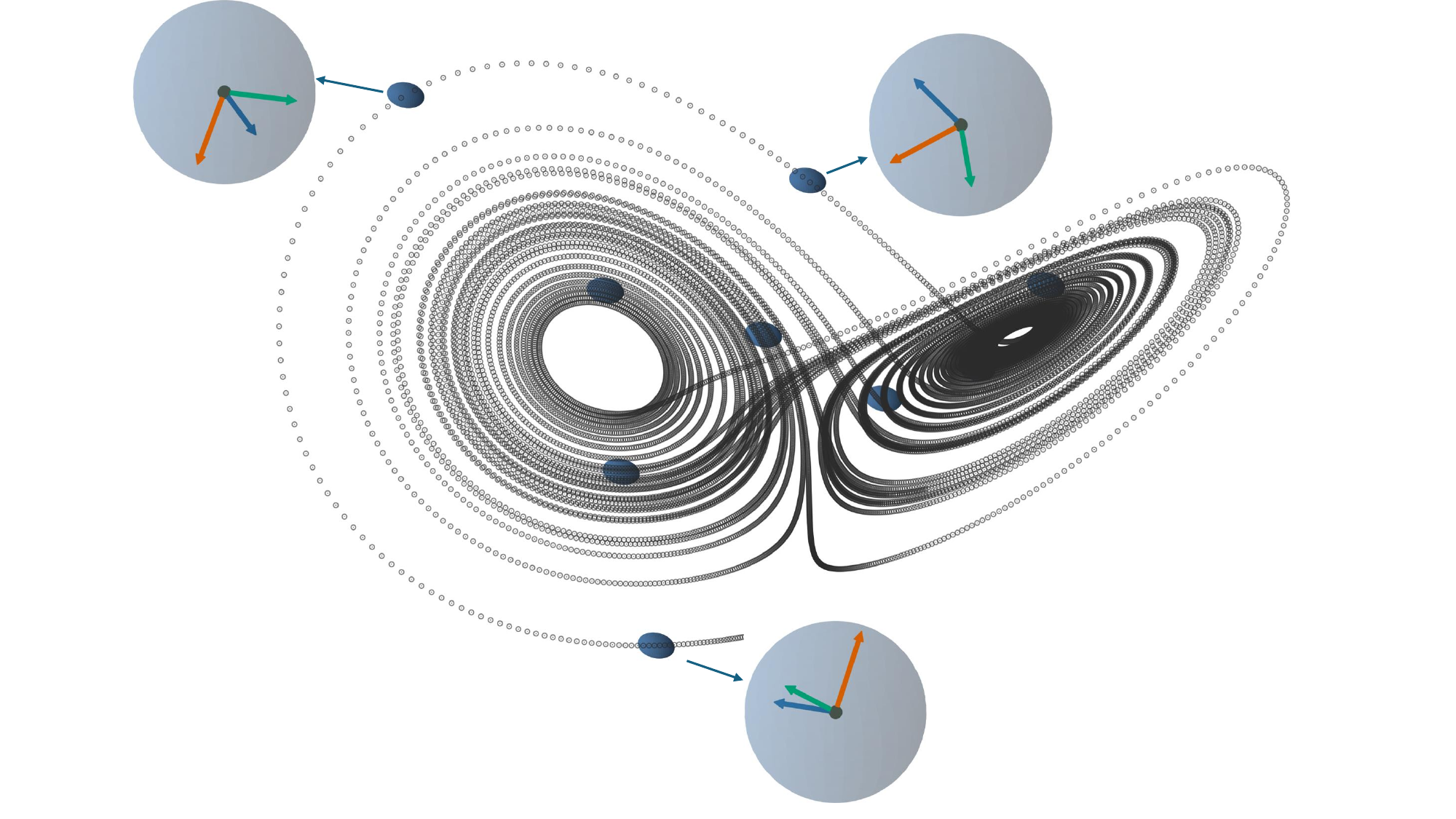}
\caption{\label{fig:Lorenz-bloch}\textit{Classical analogue of the Bloch sphere for $\jacobian_-$ dynamics.} 
Time evolution of an initial phase-space volume spanned by a complete set of
basis vectors $\ket{\delta \bm{\phi}_j}$ along the Lorenz attractor.
The associated pure states $\brho_j=\dyad{\delta \bm{\phi}_j}{\delta \bm{\phi}_j}$
define a unit 2-sphere whose volume remains invariant under the $\jacobian_-$
dynamics at each phase-space point on the attractor.
Starting from a sphere at time $t_0$ constructed from three mutually orthogonal
basis vectors, we evolve the sphere using $\jacobian_-$ and display it at several
representative phase-space locations (small spheres along the trajectory).
For three selected locations, enlarged views are provided to show explicitly the
three orthonormal basis vectors within the sphere. Distinct colors identify the
different basis directions. At any point, the basis vectors determine the local volume variation rate $\Lambda = \sum_{j=1}^N \Tr(\stability_+ \brho_j)$.}

\end{figure}

To illustrate the volume-preserving dynamics generated by $\jacobian_-$, we consider the Lorenz-Fetter model~\cite{Lorenz63}.
In Fig.~\ref{fig:Lorenz-bloch}, we select a complete orthonormal set of perturbation vectors on the attractor and evolve them using $\jacobian_-$.
These vectors define a unit 2-sphere in tangent space whose volume remains invariant as it is transported along the trajectory.
Under $\jacobian_-$ dynamics the basis vectors rotate while preserving their mutual orthogonality, unit norm, and the phase space volume they span.
In contrast, the underlying chaotic trajectory continues to exhibit the usual sensitivity associated with unstable manifolds.
Enlarged spheres display the orthonormal basis vectors at selected phase-space points.

In general, at every phase point in an $(n+1)$-dimensional phase space, there is a unit $n$-sphere of volume $\pi^{n/2}/\Gamma(n/2+1)$, where $\Gamma$ is the gamma function. This spherical phase volume remains conserved for the $\jacobian_-$ dynamics for any deterministic system. 
This sphere is a classical analogue of the Bloch sphere~\cite{nielsen2010quantum}.
Geometrically, the classical Bloch sphere is centered at the phase space point and any point on the surface of the sphere represents a tangent vector at that phase point. 
The surface of the classical Bloch sphere contains phase space points local to the reference point at the center. 
In this picture, the local expansion/contraction rates at any point $\ket{\delta \yu}$ on the surface of this invariant phase volume can be computed as $\bra{\delta\yu}\stability_+\ket{\delta\yu}=\Tr(\stability_+\brho)$. 
Similar computations for pure states in the complete set of basis vectors leads to the entire spectrum of instantaneous Lyapunov exponents.
We discuss these basis states below.

\begin{table*}[ht]
	\caption{For $\widetilde{\jacobian}$ dynamics: Equation of motion and instantaneous Lyapunov exponent for the pure state $\brho_i = \dyad{\delta\bphi_i}{\delta\bphi_i}$ in the eigenbases of $\stability_+$, $\stability_-$, and $\stability$, for Hamiltonian and dissipative dynamics.}
	\label{table:summary}
		\begin{tabular}{cccc} 
		\hline
		\hline
		& $\stability_+$ & $\stability_-$ & $\stability$ \\ 
		\hline
		\multirow{2}{6.5em}{\textit{Hamiltonian}} & $d_t\brho_i= [\stabilityH_-,\brho_i]$ & $d_t\brho_i=\{\stabilityH_+,\brho_i\}$& $d_t\brho_i=0$\\
		& $r^{\stabilityH_+}_i = \operatorname{Eig}\stabilityH_{+,i}$ & $r_i^{\stabilityH_-} = 0$ & $r_i^{\stabilityH}=\Re(\operatorname{Eig}\stabilityH_i)$ \\
		\hline
		\multirow{2}{6em}{\textit{Dissipative}} & $d_t\brho_i= [\stability_-,\brho_i]$ & $d_t\brho_i=\{\stability_+,\brho_i\}-2r^{\stability_-}_i\brho_i$& $d_t\brho_i=0$ \\
		& $r_i^{\stability_+} = \operatorname{Eig}\stability_{+,i}$ & $r_i^{\stability_-} = \Tr(\stability_+\brho_i)$ & $r^{\stability}_i=\Re(\operatorname{Eig}\stability_i)$\\
		\hline
		\hline
	\end{tabular}
\end{table*}

\section{Basis states and their stretching rates}\label{sec:basis}

Local rates for stretching and contractions are important to characterize the evolution and determine the stability of dynamical systems.
The sum of these rates is the local phase space contraction rate, which contributes to the local Gibbs entropy flow rate~\cite{Andrey1985a, Andrey1985b,TucLiuGioMar2001, Patra2016, DasGreen-spectral-bounds}.
These local or instantaneous Lyapunov exponents depend on local directions in phase space.
While asymptotically the local tangent vectors converge to the most expanding directions, they rapidly fluctuate on short times~\cite{Eckmann1985Ergodic, hoover2001fluctuations}. 
The set of these rates computed for the complete set of orthogonal directions provides the spectrum of finite-time or local Lyapunov exponents. 
In our density matrix theory, each of these rates is an expectation value of $\stability_+$ over a distinct pure state.

To compute the full spectrum of these exponents, we can choose a set of basis vectors in the tangent space $T\mathcal{M}$ attached to a point in phase space $\mathcal{M}$.
Consider a pure state formed by the basis vector $\ket{\delta\bm{\phi}_i}\in T\mathcal{M}$:
\begin{align}
\brho_i(t) & = \dyad{\delta\bm{\phi}_i(t)}{\delta\bm{\phi}_i(t)}.
\end{align}
There are uncountably many sets of linearly independent vectors that span an $n$-dimensional phase space.
A simple choice is the Cartesian coordinates fixed on the given trajectory, $\ket{\delta \bm{\phi}_i} = \delta_{ij}\ket{\mathbb{1}}$, where $\delta_{ij}$ is the Kronecker delta, $i=1,\ldots,n$, and $\ket{\mathbb{1}}$ represents an $n\times 1$ unit matrix.
The instantaneous Lyapunov exponents for these basis states are the diagonal elements of the stability matrix $\stability$:
\begin{align}
r_i = \bra{\delta \bm{\phi}_i}\stability\ket{\delta \bm{\phi}_i} = \Tr(\stability_+\brho_i) = (\boldsymbol{A})_{ii}.
\end{align}
Other coordinate systems can be chosen to represent the basis states, which may be time-independent or time-dependent and comoving with the phase point.

The time evolution of a set of complete basis states at an initial time $t_0$ is governed by the operator $\widetilde{\jacobian}$ (or $\widetilde{\jacobian}_+$ or $\jacobian_-$).
Alternatively, a set of basis states can be chosen at each phase space point (we give some examples below).
	The basis vectors can be used to write the corresponding maximally mixed state at time $t$ using Eq.~\ref{eq:maximally_mixed}.
One can construct another mixed state at the same time $t$, say $\brho^d$ using a set of basis states defined by the original dynamics.
These two states $\brho$ and $\brho^d$ are related by a similarity transformation
\begin{align}
	\brho^d = \boldsymbol{P}\brho\boldsymbol{P}^{-1}, 
\end{align}
where  $\boldsymbol P$ is a change of basis matrix. 
Therefore, all such maximally mixed states are \emph{similar} to each other and give the local phase space dissipation rate when $\stability_+$ is averaged over any of them (see Eq.~\ref{eq:dissipation_rate}).

As shown in Appendix~\ref{SI:comparison}, there is a relationship between the antisymmetric evolution and the continuous QR/Gram-Schmidt procedure.
The QR dynamics combines the physical rotation generated by $\stability_-$ with an additional constraint-induced rotation arising from $\stability_+$, whereas the present splitting isolates the intrinsic rotational component.
This distinction emphasizes that the $\jacobian_-$ evolution captures only the physical rotation of tangent vectors by decomposing tangent-space dynamics at the level of the generator, while QR incorporates both rotation and stretching along with frame adjustments to maintain orthonormality.

Next we analyze several flow-dependent basis vectors to demonstrate the computation of the entire spectrum of  instantaneous Lyapunov exponents in this framework.

\smallskip
\noindent\textit{a. $\stability_{+}$ basis.--} Eigenvalues of $\stability_{+}$, $\sigma^{\stability_+}_{i}$, satisfy
$\stability_{+}\brho_i = \sigma^{\stability_+}_{i}\brho_i$.
The instantaneous Lyapunov exponents are tangent space averages:
\begin{align}
r_i^{\stability_+} &= \Tr(\stability_+\brho_i) = \sigma^{\stability_+}_{i}.
\end{align}
Since $\stability_+$ is symmetric, its eigenvalues are real.

	When $\brho_i$ is evaluated in the eigenbasis of the symmetric operator $\stability_+$, the anti-commutator term in the equation of motion Eq.~\ref{eq:EOM_rho} vanishes. As a result, for the $\widetilde{\jacobian}$ dynamics the evolution of $\brho_i$ reduces to  (see Appendix~\ref{SI:A-basis-EOM}):
\begin{equation}
\frac{d\brho_i}{dt} = [\stability_-,\brho_i].
\end{equation}
The same equation of motion holds for the $\jacobian_-$ dynamics. However, for $\widetilde{\jacobian}_+$, the corresponding equation reduces to $d_t\brho_i = \mathbb{0}$.  This means that in the $\stability_+$ eigenbasis, the operator $\widetilde{\jacobian}_+$ generated dynamics preserves the basis state $\brho_i$.

\smallskip

A discussion on other sets of possible basis states needs a generalization of the formalism to finite-dimensional complex vector spaces.
Consider a generalized basis state given by a complex vector $\ket{\delta \yu} \in \mathbb{C}^n$ and a Hermitian density matrix $\brho$.
In this case, we can update our use of Dirac's notation to now represent a finite-dimensional column (row) vector with the ket (complex conjugate bra).
However, the operators $\stability$ and $\stability_\pm$ are still given by real matrices.
The local Lyapunov exponents are $\Tr(\stability_+\brho)$ also real for a Hermitian $\brho$ as we will see next.

\smallskip

\noindent\textit{b. $\stability_-$ basis.--}
The eigenvectors of the anti-symmetric matrix $\stability_-$ form a complete basis but are not necessarily orthogonal.
The eigenvalues of $\stability_-$ are purely imaginary, so $[\stability_-,\brho_i] = \stability_-\brho_i - \brho_i\stability_- = \mathbb{0}.$ Eq.~\ref{eq:EOM_rho} then becomes
\begin{equation}
\frac{d\brho_i}{dt} = \{\stability_+,\brho_i\} -  2r_i^{\stability_-}\brho_i,
\end{equation}
with the instantaneous Lyapunov exponent $r_i^{\stability_-}$ in the eigenbasis of $\stability_-$. The generalized Liouville equation for the maximally mixed state $\brho$ in this basis is then
\begin{equation}
\frac{d|\brho|}{dt} = \bigl\rvert\{\stability_+,\brho\} - 2r\brho\bigr\rvert = 0.
\end{equation}

\smallskip
\noindent\textit{c. $\stability$ basis.--} The eigenvectors of $\stability$ form a complete set of basis but not mutually orthogonal. If the eigenvalues of $\stability$ are denoted by $\sigma^{\stability}_i$, then the instantaneous Lyapunov exponents for these basis states are derived as follows
\begin{align}
r_i^{\stability} &= \Tr(\stability_+\brho_i) = \frac{1}{2}\Tr(\stability\brho_i + \stability^\top\brho_i) \nonumber\\
&= \frac{1}{2}(\sigma^{\stability}_i + \sigma_i^{\dagger\stability}) =\Re(\sigma^{\stability}_i)\label{eq:ILE_A},
\end{align}
where $\dagger$ denotes the complex conjugate and $\Re(.)$ gives the real part. 

\begin{table*}[ht]
	\caption{Instantaneous Lyapunov exponents for the linear and damped harmonic oscillator in the eigenvector basis of the stability matrix $\stability$ and its (anti)symmetric parts. The harmonic oscillator with frequency $\omega$ has the Hamiltonian: $H = p^2 + \omega^2q$; in this case, $\{\Tr \brho,H\} = 0$. The equations of motion for the damped harmonic oscillator with the dissipation parameter $\gamma$ are $\dot{q} = p$ and $\dot{p} = -\omega^2q - \gamma p$.}
	\label{table:HOsummary}
	
	\begin{tabular}{cccc} 
		\hline
		\hline
		Harmonic oscillator & $\stability_+$ & $\stability_-$ & $\stability$\\ 
		\hline
		\vspace{2pt}
		\textit{Linear} & $r_i^{\stabilityH_+} = \pm (1-\omega^2)/2$&
		\vspace{2pt}
		$r_i^{\stabilityH_-} = 0$&
		$r_i^{\stabilityH} = 0$\\[2pt]
		\textit{Damped} & $r_i^{\stability_+} = \frac{1}{2}(-\gamma \pm \sqrt{\gamma^2 + (1-\omega^2)^2})$ &
		$r_i^{\stability_-} = -\frac{1}{2}\gamma$ &
		$r_i^{\stability} = \frac{1}{2}(-\gamma \pm \Re (\sqrt{\gamma^2 - 4\omega^2}))$\\[2pt]
		\hline
		\hline
	\end{tabular}
\end{table*}

\begin{figure*}[!t]
		\includegraphics[width=0.9\textwidth]{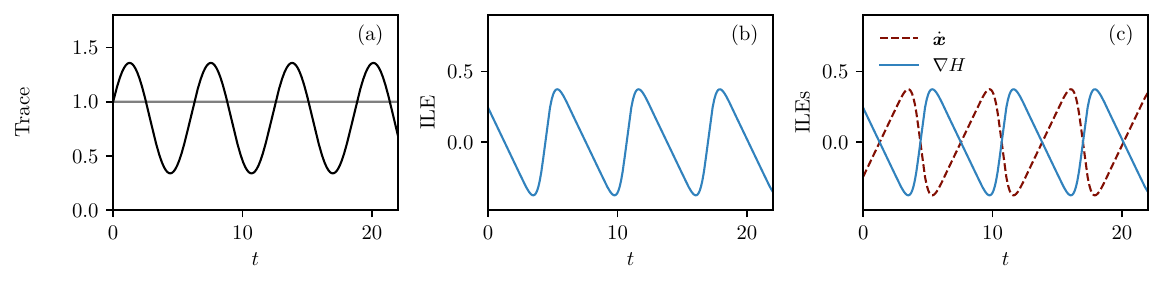}
		\caption{\label{fig:SI_HO} 
		(a) Linear harmonic oscillator ($\omega=0.5$): $\Tr\bxi$ (black) and $\Tr\brho$ (gray) as a function of time.
(b) Time evolution of the instantaneous Lyapunov exponent (ILE) for an arbitrary pure state.
(c) ILEs in the conjugate tangent-space directions $\dot{\ex}$ (red, dashed) and $\grad H$ (blue, solid).
}
\end{figure*}

\subsection*{Hamiltonian systems}

\begin{figure*}[!t]
		\includegraphics[width=0.9\textwidth]{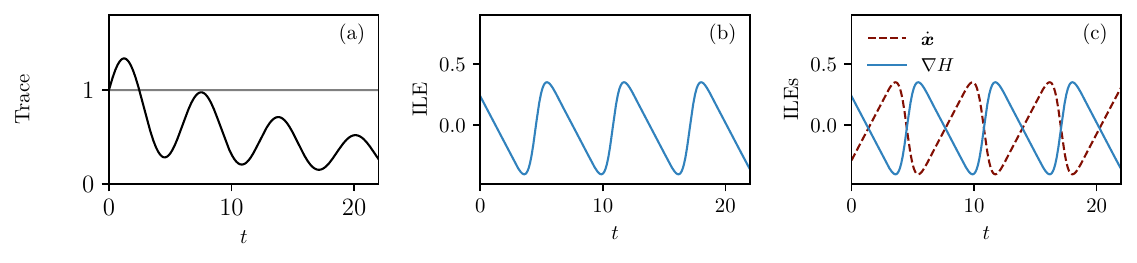}
		\caption{\label{fig:SI_DHO}
Damped harmonic oscillator ($\omega=0.5$, $\gamma=0.05$):
(a) Time evolution of $\Tr\bxi$ (black) and $\Tr\brho$ (gray).
(b) Instantaneous Lyapunov exponent (ILE) for an arbitrary pure state.
(c) ILEs in the conjugate tangent-space directions $\dot{\ex}$ (red, dashed) and $\grad H$ (blue, solid).
}
\end{figure*}

Given the Liouville's theorem is a statement about volume conservation, a natural example to consider is Hamiltonian dynamics.
For Hamiltonian systems, the components of a Lyapunov vector are the first variations of position and momentum $(\delta \boldsymbol{q}, \delta \boldsymbol{p})$ and the stability matrix $\stability$ becomes:
\begin{eqnarray}\nonumber
\stabilityH&=&\left(
\begin{array}{cc}
\partial_\bmQ\partial_\bmP\Hamiltonian & \partial_\bmP^2\Hamiltonian\\
-\partial_\bmQ^2\Hamiltonian & -\partial_\bmP\partial_\bmQ\Hamiltonian\\
\end{array}\right)=\left(
\begin{array}{cc}
\mathbb{0} & \partial_\bmP^2\Hamiltonian\\
-\partial_\bmQ^2\Hamiltonian & \mathbb{0}\\
\end{array}\right).
\end{eqnarray}
The trace of the Jacobian gives the traditional form of Liouville's theorem.
In the last equality, we assume the Hamiltonian is $H(\boldsymbol{q},\boldsymbol{p}) = T(\boldsymbol{p}) + V(\boldsymbol{q})$ and the stability matrix is the product of its Hessian and the symplectic matrix:
\begin{equation*}
\bm{\Omega}=\left(
\begin{array}{cc}
\mathbb{0} & \mathbb{1}\\
-\mathbb{1} & \mathbb{0}\\
\end{array}\right).
\end{equation*}
The identity matrices $\mathbb{I}$ are $n/2\times n/2$.

The symmetric and anti-symmetric parts of the stability matrix are
\begin{align}
\stabilityH_+ &= \frac{1}{2}(\partial_\bmP^2\Hamiltonian-\partial_\bmQ^2\Hamiltonian)\left(
\begin{array}{cc}
\mathbb{0} & \mathbb{1}\\
\mathbb{1} & \mathbb{0}\\
\end{array}\right) \\
\stabilityH_- &= \frac{1}{2}(\partial_\bmP^2\Hamiltonian+\partial_\bmQ^2\Hamiltonian)\left(
\begin{array}{cc}
\mathbb{0} & \mathbb{1}\\
-\mathbb{1} & \mathbb{0}\\
\end{array}\right),
\end{align}
respectively. Equation~\ref{eq:EOM_rho} then becomes
\begin{align}
\frac{d\brho}{dt} &= \{\stabilityH_+,\brho\} + [\stabilityH_-,\brho] - 2r\brho.
\label{Liouville_Lyap_eq_H_dynamics}
\end{align}
Since the traditional forms for Liouville's theorem/equation are specific to Hamiltonian dynamics, we consider a set of eigenbases for this special case.

\smallskip
\noindent\textit{a. $\stabilityH_{+}$ basis.--} 
The eigenvectors of $\stabilityH_{+}$ are a complete set of orthonormal basis vectors in an $n$-dimensional phase space, with a general form:
\begin{align}
\ket{\delta \bphi_{qj}} = \frac{1}{\sqrt{2}}(\delta_{jk}\ket{\mathbb{1}}+\delta_{j'k'}\ket{\mathbb{1}}),\nonumber \\
\ket{\delta \bphi_{pj}} = \frac{1}{\sqrt{2}}(\delta_{jk}\ket{\mathbb{1}}-\delta_{j'k'}\ket{\mathbb{1}}),
\end{align}
where $\delta_{jk}$ and $\delta_{j'k'}$ are Kronecker deltas: $\delta_{jk} = 1$ for $j=k$ and zero otherwise. Indices $j'$ and $j$ are related: $j' = n/2+j$, where $j$ runs from $1$ to $n/2$. The same holds for the indices $k$ and $k'$.

For example, in the case of a 2D-Hamiltonian system, these basis vectors become
\begin{align}
\ket{\delta \bm{\phi}_1} &= \frac{1}{\sqrt{2}}\begin{pmatrix}
1\\ 
1
\end{pmatrix},\quad\ket{\delta\bm{\phi}_2} =\frac{1}{\sqrt{2}}\begin{pmatrix}
1\\ 
-1\\
\end{pmatrix},
\end{align} 
which define the basis states:
\begin{align}
\brho_1 &=  \frac{1}{2}
\begin{pmatrix}
1 &	1\\ 
1 &	1
\end{pmatrix},\quad
\brho_2 =  \frac{1}{2}\begin{pmatrix}
1 &	-1\\ 
-1 & 1\\
\end{pmatrix}.
\end{align}
For these states, the instantaneous Lyapunov exponents are the eigenvalues of $\stabilityH_{+}$,
\begin{align}\label{eq:ILE_symA}
r_k^{\stabilityH_+} = \pm  \frac{1}{2}(\partial_\bmP^2\Hamiltonian-\partial_\bmQ^2\Hamiltonian)=  \pm  \frac{1}{2}\,\grad\cdot 
\begin{pmatrix} \dot p \\ \dot q \end{pmatrix}.
\end{align}

\smallskip
\noindent\textit{b. $\stabilityH_{-}$ basis.--} The eigenvectors of $\stabilityH_-$ are also a complete set of orthonormal basis vectors,
\begin{align}
\ket{\delta \bphi_{qj}} = \frac{1}{\sqrt{2}}(\delta_{jk}\ket{\mathbb{1}}+i\delta_{j'k'}\ket{\mathbb{1}}),\nonumber \\
\ket{\delta \bphi_{pj}} = \frac{1}{\sqrt{2}}(\delta_{jk}\ket{\mathbb{1}}-i\delta_{j'k'}\ket{\mathbb{1}}),
\end{align}
where $\delta_{jk}$ and $\delta_{j'k'}$ are Kronecker deltas with $j' = n/2+j$.

Again, for a 2D-Hamiltonian system, these states become
\begin{align}
\ket{\delta \bphi_1} &= \frac{1}{\sqrt{2}}\begin{pmatrix}
1\\ 
i
\end{pmatrix}, \quad
\ket{\delta \bphi_2}=\frac{1}{\sqrt{2}}\begin{pmatrix}
1\\ 
-i
\end{pmatrix},
\end{align}
with purely imaginary eigenvalues $\pm i(\partial_\bmP^2\Hamiltonian+\partial_\bmQ^2\Hamiltonian) $.
The corresponding basis states are:
\begin{align}\label{eq:basis_states_A_anti}
\brho_1 &=  \frac{1}{2}
\begin{pmatrix}
1 & -i\\ 
i &1
\end{pmatrix},\quad
\brho_2 =  \frac{1}{2}\begin{pmatrix}
1 &	i\\ 
-i & 1\\
\end{pmatrix}.
\end{align}
For these basis states,
\begin{align}
r_k^{\stability_-} & = \Tr(\stabilityH_+\brho_k) = 0,
\end{align} 
the instantaneous Lyapunov exponents in the $\stabilityH_-$ basis always vanishes for Hamiltonian systems.
This fact further simplifies Eq.~\ref{eq:EOM_rho} to,
\begin{equation}
\frac{d\brho_k}{dt} = \{\stability_+,\brho_k\},
\end{equation}
and equation of motion that depends only on the classical anti-commutator.

\noindent\textit{c. $\stabilityH$ basis.--} From Eq.~\ref{eq:ILE_A}, we know that the instantaneous Lyapunov exponents in this basis are the real part of the eigenvalues of $\stabilityH$,
\begin{align}
\lambda^{\stabilityH}_k &= \pm i \sqrt{\partial^2_{p_k}\Hamiltonian\,\partial^2_{q_k} H}.
\end{align}
So, for non-zero exponents, we must have $\partial^2_{p_k}\Hamiltonian\,\partial^2_{q_k} H<0$. Equation~\ref{Liouville_Lyap_eq_H_dynamics} in this basis becomes $d_t\brho_k = \mathbb{0}$.

\smallskip
\noindent\textit{d. Tangent pure states.--} In Ref.~\onlinecite{dasDensityMatrixFormulation2022}, we introduced in  a set of conjugate pure states in the tangent space for a 2D-Hamiltonian system:
\begin{align}
\|\dot \ex\|^2\brho_{\dot \ex} &=
\begin{pmatrix}
\dot q^2 & \dot p\dot q \\[2pt]
\dot p\dot q & \dot p^2\\ 
\end{pmatrix},
\|\grad\Hamiltonian\|^2\brho_{\grad\Hamiltonian} 
&=
\begin{pmatrix}
\dot p^2 & -\dot p\dot q\\
-\dot p\dot q & \dot q^2
\end{pmatrix}.\label{eq:tangent_pure_states}
\end{align}
These states are formed from the outer product of the unit tangent vectors, $\ket{\delta\bphi_{\dot{\ex}}} = \|\dot{\ex}\|^{-1}(\dot q,\dot p)^\top$ and $\ket{\delta\bphi_{\grad{\Hamiltonian}}} = \|\grad\Hamiltonian\|^{-1}(-\dot{p},\dot{q})^\top$,
where $\|\dot \ex\|^2 = \|\grad\Hamiltonian\|^2 = \dot{p}^2 + \dot{q}^2$. 
These are, on a constant energy manifold, two conjugate tangent directions: the phase velocity $\dot \ex$ and the gradient of the Hamiltonian $\grad\Hamiltonian$, related by $\dot \ex=\bm{\Omega} \grad\Hamiltonian$ through the symplectic matrix $\bm{\Omega}$. These are orthogonal to each other, $\grad\Hamiltonian\cdot \bm{\Omega} \grad \Hamiltonian = 0$, and have equal magnitude $\|\dot \ex\|=\|\grad \Hamiltonian\|$ through Hamilton's equations.

Consider conjugate pure states in the tangent space for Hamiltonian dynamics. For a 2D-Hamiltonian system,
\begin{align}
\|\dot \ex\|^2\brho_{\dot \ex} &= 
\begin{pmatrix}
(\partial_p \Hamiltonian)^2 & -\partial_p \Hamiltonian\partial_q \Hamiltonian\\[2pt]
-\partial_q \Hamiltonian\partial_p \Hamiltonian & (\partial_q \Hamiltonian)^2\\ 
\end{pmatrix}=
\begin{pmatrix}
\dot q^2 & \dot p\dot q \\[2pt]
\dot p\dot q & \dot p^2\\ 
\end{pmatrix}, \nonumber 
\end{align}
\begin{align}
\|\grad\Hamiltonian\|^2\brho_{\grad\Hamiltonian} &= 
\begin{pmatrix}
(\partial_q \Hamiltonian)^2 & \partial_q \Hamiltonian\partial_p \Hamiltonian\\[2pt]
\partial_p H\partial_q \Hamiltonian & (\partial_p \Hamiltonian)^2\\ 
\end{pmatrix}=
\begin{pmatrix}
\dot p^2 & -\dot p\dot q\\
-\dot p\dot q & \dot q^2
\end{pmatrix}. \nonumber
\end{align}
For an $n$-dimensional system, these pure states are formed by the basis vectors
\begin{align}
	\ket{\delta \bphi_{\dot{\ex}_j}} &= \frac{1}{\|\dot{\ex}_j\|}(\dot q_j\delta_{jk}\ket{\mathbb{1}}+\dot p_j\delta_{j'k'}\ket{\mathbb{1}}),\nonumber \\
\ket{\delta \bphi_{\boldsymbol{\nabla} H_j}} &=\frac{1}{\|\boldsymbol{\nabla} H_j\|}(-\dot p_j\delta_{jk}\ket{\mathbb{1}}+\dot q_j\delta_{j'k'}\ket{\mathbb{1}}),
\end{align}
with Kronecker deltas $\delta_{jk}$ and $\delta_{j'k'}$ and $j' = n/2+j$.

For these time-dependent basis states, the instantaneous Lyapunov exponents are
\begin{align}\label{eq:ILE_tangent}
r_k &= \pm  \frac{\dot q_k\dot p_k}{||\dot \ex_k||^2}\grad\cdot 
\begin{pmatrix} \dot p_k \\ \dot q_k \end{pmatrix} = \pm \Tr(\brho_{\dot \ex_k}\stabilityH_+).
\end{align}
Table~\ref{table:summary} shows the summary of results in these bases for Hamiltonian and dissipative systems.

\section{Case studies}\label{sec:case_examples}

\noindent\textit{a. Linear harmonic oscillator.--} We consider a one-dimensional harmonic oscillator with unit mass and frequency $\omega$. 
Using dimensionless variables---where position is scaled by the characteristic length $1/\sqrt{\omega}$, momentum by $\sqrt{\omega}$, and energy by $\omega$---the Hamiltonian takes the form $H = \frac{1}{2}(p^2 + \omega^2 q^2)$. The corresponding equations of motion are $\dot{q} = p$ and $\dot{p} = -\omega^2 q$, which yield a time-independent stability matrix:
\begin{align}
\stabilityH &= \left(
\begin{array}{cc}
0 & 1\\
-\omega^2 & 0\\
\end{array}\right),
\end{align}
with symmetric and anti-symmetric parts
\begin{align}
\stabilityH_+&= \frac{1}{2}\left(
\begin{array}{cc}
0 & 1-\omega^2\\
1-\omega^2 & 0\\
\end{array}\right), \nonumber\\
\stabilityH_-&= \frac{1}{2}\left(
\begin{array}{cc}
0 & 1+\omega^2 \\
-1-\omega^2 & 0\\
\end{array}\right).
\end{align}
In the basis of $\stabilityH$ and $\stabilityH_-$, the instantaneous Lyapunov exponents vanish.
Only the instantaneous Lyapunov exponents in the $\stabilityH_+$ eigenvector basis is nonzero, Table~\ref{table:HOsummary}.
Because the dynamics are Hamiltonian, there is a conjugate pair of exponents with a magnitude that depends on the oscillator frequency $\omega$.
Fig.~\ref{fig:SI_HO}(a) shows the time evolution of $\Tr\bxi$ for a pure state formed from a random unit perturbation vector $\ket{\delta \ex}$ on a sampled trajectory: $\bxi = \dyad{\delta \ex}{\delta \ex}$.
Because this density matrix is unnormalized, $\Tr \bxi \neq 1$.
Its normalized counterpart $\brho$ has trace 1 throughout the evolution (indicated by horizontal gray line $\Tr \brho = 1$) in Fig.~\ref{fig:SI_HO}(a).
For this pure state, the instantaneous Lyapunov exponent is shown in ~\ref{fig:SI_HO}(b).

The instantaneous Lyapunov exponents in the conjugate tangent space directions $\dot\ex$ and $\grad H$ for the linear harmonic oscillator are also a conjugate pair. From Eq.~\ref{eq:tangent_pure_states}, the pure states are:
\begin{align}
\brho_1 &=  \|\dot \ex\|^{-2}
\begin{pmatrix}
p^2 &	-\omega^2pq\\ 
-\omega^2pq &	\omega^4q^2
\end{pmatrix}\nonumber \\
\brho_2 &=  \|\grad\Hamiltonian\|^{-2}\begin{pmatrix}
\omega^4q^2 &	\omega^2pq\\ 
\omega^2pq & p^2\\
\end{pmatrix},
\end{align}
where $\|\dot \ex\|^{2} = \|\grad\Hamiltonian\|^2 = p^2 + \omega^4q^2$. The instantaneous Lyapunov exponents for these states,
\begin{align}\label{eq:ILE_HO}
r_{1,2} & = \Tr(\stabilityH_{+}\brho_{1,2})= \pm (1-\omega^2)\omega^2\frac{pq}{p^2+\omega^4q^2},
\end{align}
are functions of time, as shown in Fig.~\ref{fig:SI_HO}(c) using a frequency of $\omega = 0.5$.
In the eigenbasis of $\stabilityH_+$, the instantaneous Lyapunov exponents are (from Appendix~\ref{SI:evol-mat} Eq.~\ref{eq:ILE_symA}): $r^{\stabilityH_+}_{1,2} = \pm \frac{1}{2}(1-\omega^2)$.
Therefore, the instantaneous Lyapunov exponents in these special directions are the eigenvalues of $\stabilityH_+$ scaled by a function of the state. 

These results can be generalized to higher dimensions, as we show for the H\'enon-Heiles system in Subsection~\ref{subsec:HH-model} C. 

\noindent\textit{b. Damped harmonic oscillator.--}
 For the damped harmonic oscillator and a velocity dependent dissipation parameterized by $\gamma$, the equations of motion are:
\begin{align}\label{eq:DHO}
\dot{q} &= p  \nonumber\\
\dot{p} &= -\omega^2 q - \gamma p.
\end{align} 
Conservative dynamics are recovered for $\gamma = 0$.
The stability matrix of this system,
\begin{align}
\stability =\left(
\begin{array}{cc}
0 & 1\\
-\omega^2 & -\gamma\\
\end{array}\right),
\end{align}
has symmetric and anti-symmetric parts:
\begin{align}\nonumber
\stability_+ &=\frac{1}{2}\left(
\begin{array}{cc}
0 & 1-\omega^2\\
1-\omega^2 & -2\gamma\\
\end{array}\right), \\
\stability_- &=\frac{1}{2}\left(
\begin{array}{cc}
0 & 1+\omega^2\\
-1-\omega^2 & 0\\
\end{array}\right). 
\end{align}
For a pure state (unnormalized) $\Tr \bxi$ constructed from a random tangent vector $\ket{\delta \ex}$, Fig.~\ref{fig:SI_DHO}(a) shows the time evolution of trace $\Tr\bxi$ and the trace of its normalized version $\Tr \brho$.
Figure~\ref{fig:SI_DHO}(b) shows the corresponding instantaneous Lyapunov exponent.

The instantaneous Lyapunov exponents in the conjugate tangent space directions  for the damped harmonic oscillator are found, as before, from the pure states in Eq.~\ref{eq:tangent_pure_states}:
\begin{align}
\brho_1 &=  \|\dot \ex\|^{-2}
\begin{pmatrix}
p^2 &	-(\omega^2q +\gamma p)p\\ 
-(\omega^2q +\gamma p)p &	(\omega^2q + \gamma p)^2
\end{pmatrix}\nonumber \\
\brho_2 &=  \|\grad\Hamiltonian\|^{-2}\begin{pmatrix}
(\omega^2q + \gamma p)^2 &	(\omega^2q +\gamma p)p\\ 
(\omega^2q +\gamma p)p & p^2\\
\end{pmatrix},
\end{align}
where $\|\dot \ex\|^{2} = \|\grad\Hamiltonian\|^2 = p^2 + (\omega^2q + \gamma p)^2$. The instantaneous Lyapunov exponent for these states,
\begin{align}
r_{1} & = \Tr(\stability_{+}\brho_{1})\nonumber\\
&= \frac{-(\omega^2q + \gamma p)}{p^2+(\omega^2q + \gamma p)^2}\Big[(1-\omega^2)p+ \gamma(\omega^2q+\gamma p)\Big], \nonumber\\[5pt]
r_{2} & = \Tr(\stability_{+}\brho_{2})\nonumber\\
& = \frac{\omega^2p}{p^2+(\omega^2q + \gamma p)^2}\Big[(1-\omega^2)q - \gamma p\Big].
\end{align}
are shown in Fig.~\ref{fig:SI_DHO}(c) using a frequency of $\omega = 0.5$ and $\gamma=0.05$.
Equation~\ref{eq:ILE_HO} is recovered when $\gamma = 0$.

\begin{figure*}[!t]
	\includegraphics[width=0.9\textwidth]{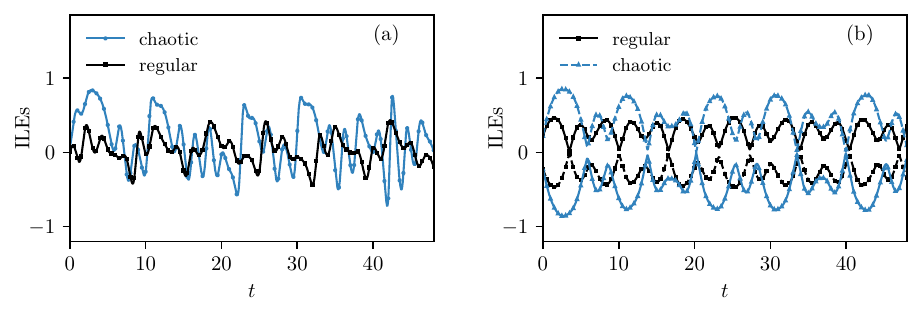}	
	\caption{\label{fig:SI_HH}
The H\'enon--Heiles model: a regular orbit at energy $E=0.0833$ and a chaotic orbit at energy $E=0.1677$.
(a) Instantaneous Lyapunov exponent (ILE) for representative pure states along the two trajectories.
(b) ILEs in the $\stabilityH_+$ basis for the regular (black, squares) and chaotic (blue, triangles) orbits.
In panel (b), two ILEs are shown for each orbit, corresponding to a Hamiltonian pair $\lambda$ and $-\lambda$; markers distinguish regular and chaotic dynamics.}

\end{figure*}

We can also compute these instantaneous Lyapunov exponents in the other eigenbases. For instance, in the $\stability_+$-basis, its eigenvalues are:
\begin{align}
r^{\stability_+}_{1,2} &= \frac{1}{2}(-\gamma \pm \sqrt{\gamma^2 + (1-\omega^2)^2}).
\end{align}
In the $\stability$ eigenbasis, these are $r_i^{\stability} = \Re(\lambda^\stability_{i})$ (using Eq.~\ref{eq:ILE_A}),
\begin{align}
r_i^{\stability} = \Re \lambda^\stability_i = \Re \frac{1}{2}(-\gamma \pm \sqrt{\gamma^2 - 4\omega^2}),
\end{align} 
where $\lambda^\stability_i$ indicates the eigenvalue of $\stability$. 
The real part of $\lambda^\stability_i$ gives the instantaneous Lyapunov exponent in the $\stability$ basis. These exponents are unequal when $\gamma^2>4\omega^2$. 
In the $\stability_-$ basis, we can compute the instantaneous Lyapunov exponents as an expectation value over the normalized density matrix, $r_i^{\stability_-} = \Tr (\stability_-\brho_i)$, where each $\brho_i$ is the eigenvectors of $\stability_-$ (see Eq.~\ref{eq:basis_states_A_anti}).
Doing so, we find the instantaneous Lyapunov exponents are directly proportional to the damping coefficient $r_i^{\stability_-} = -\gamma/2$. 

These results for a dissipative system can be also generalized to higher dimensions, as we show for the Lorenz-Fetter model later.

\noindent\textit{c. The H\'enon-Heiles system.--}
\label{subsec:HH-model}
This is a well-known model of nonlinear dynamics performed by a star around a galactic with motion restricted to a plane. It is given by the Hamiltonian~\cite{HenHeil1964}:
\begin{align}
	\Hamiltonian = \frac{1}{2}(p_x^2 + p_y^2)+ \frac{1}{2}(x^2+y^2) +\left(x^2y-\frac{y^3}{3}\right), \nonumber
\end{align}
which gives rise to a mixed phase space where regular and chaotic trajectory co-exist. The equations of motion are
\begin{align}
\dot{x} &= p_x, \quad \dot{y} = p_y,\nonumber\\
\dot{p_x} &= -x - 2xy, \quad \dot{p_y} = -y - x^2 + y^2.
\end{align}
Its stability is determined by the matrix
\begin{align}
\stabilityH =
\begin{pmatrix}
0 & 0 & 1 & 0\\
0 & 0 & 0 & 1\\
-1 - 2y & -2x & 0 & 0\\
-2x & -1+2y & 0 & 0
\end{pmatrix}.
\end{align}
Figure~\ref{fig:SI_HH}(a) shows the time evolution of a pure state described by an arbitrary unit vector on a regular and a chaotic orbit. The orbits chosen corresponds to $E=0.0833$ (regular) and $E=0.1667$ (chaotic).

The symmetric part of the stability matrix is
\begin{align}
\stabilityH_+ &=
\begin{pmatrix}
0 & 0 & -y & -x\\
0 & 0 & -x & y\\
-y & -x & 0 & 0\\
-x & y & 0 & 0
\end{pmatrix}.
\end{align}
In the eigenbasis of $\stabilityH_+$, the instantaneous Lyapunov exponents are, as usual, the eigenvalues
$r_{1,2} = \pm 2y$, and $r_{3,4} = \mp 2y$.
Figure~\ref{fig:SI_HH}(b) shows their time evolution.
The instantaneous Lyapunov exponents in the tangent space directions $\dot\ex$ and $\grad H$ are:
\begin{align}
r_{1,2} & = \pm \frac{2y(x+2xy)p_x}{p_x^2+(x+2xy)^2},
\end{align}
\begin{align}
r_{3,4} & = \mp \frac{2y(x^2-y^2+y)p_y}{p_y^2+(x^2-y^2+y)^2}.
\end{align}

\vspace*{0.5cm}

\begin{figure*}[!t]
	\includegraphics[width=0.9\textwidth]{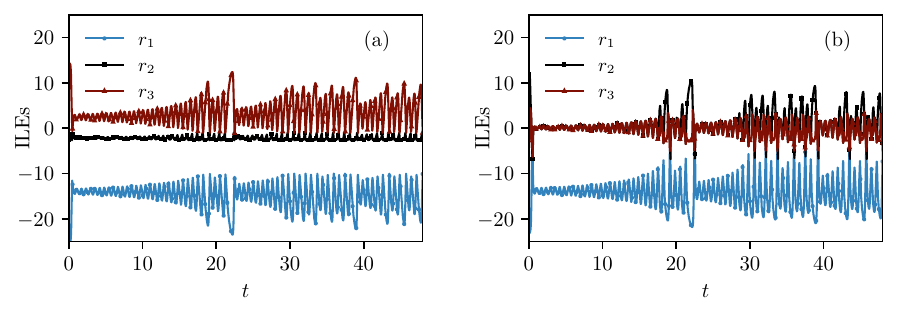}	\
	\caption{\label{fig:SI_Lorenz}
		The Lorenz-Fetter model ($\sigma = 10$, $\beta = 8/3$, $\rho = 28$): Time evolution of instantaneous Lyapunov exponents (ILEs) in (a) the $\stability_+$ basis and (b) the $\stability$ basis. Solid lines indicate the ILEs. Symbol distinguish individual exponents: in panel (a) the solid circles (in blue) indicate the smallest and the solid triangles (red) indicate the largest exponents $r_1$ and $r_3$ respectively. The other exponent, $r_2$ is shown with solid squares (black). The same symbols are used to indicate the three exponents in panel (b) in which $r_3$ is not always larger than $r_1$.
}
\end{figure*}

\noindent\textit{d. The Lorenz-Fetter system.--}
As another example, we consider the model of atmospheric convection due to Lorenz and Fetter~\cite{Lorenz63}.
The model is defined by the following systems of ordinary differential equations,
\begin{align}
\dot{x} &= \sigma(y-x), \quad\dot{y} = x(\rho - z) -y, \quad \dot{z} = xy - \beta z,
\end{align}
with the stability matrix:
\begin{align}
\stability &=\begin{pmatrix}
-\sigma & \sigma & 0\\
\rho-z & -1 & -x\\
y & x & -\beta
\end{pmatrix}.
\end{align}
The instantaneous Lyapunov exponents can be computed in any of the eigenbases using the symmetric and part of the stability matrix:
\begin{align}
\stability_+&=\begin{pmatrix}
-\sigma & \frac{1}{2}(\sigma+\rho-z) & \frac{1}{2}y\\
\frac{1}{2}(\sigma+\rho-z) & -1 & 0\\
\frac{1}{2}y & 0 & -\beta
\end{pmatrix}.
\end{align}
For the $\stability_+$ and $\stability$ bases, we plot the corresponding instantaneous Lyapunov exponents shown in Figs.~\ref{fig:SI_Lorenz}(a) and ~\ref{fig:SI_Lorenz}(b) respectively, for $\sigma = 10$, $\beta = 8/3$, $\rho = 28$.

These results demonstrate the computation of instantaneous Lyapunov exponents for both Hamiltonian and dissipative systems under the $\jacobian$ and $\jacobian_-$ dynamics, without the need for periodic re-orthonormalization of tangent vectors. The approach naturally extends, both conceptually and computationally, to higher-dimensional and many-body systems. 

\section{Conclusions}\label{sec:conclusions}

The classical density matrix theory we expand here is a useful alternative framework to generalize Liouville's theorem and equation for arbitrary deterministic systems (Hamiltonian or non-Hamiltonian). 
It leads to time invariant phase space volumes and decouples the stretching/contraction effect from the tangent space dynamics.
Here, we identified the role of symmetric and anti-symmetric part of the stability matrix $\stability$ commonly used in linearizations of dynamical systems. 
The growth or shrinking of tangent vectors is governed exclusively by $\stability_+$ while the anti-symmetric part $\stability_-$ generates only rotations in the tangent space. 
Based on this partition, we introduced two time evolution operators $\widetilde{\jacobian}$ and $\jacobian_-$, both of which preserve the norm of a perturbation vector evolving along a phase space trajectory. 

	Both time-evolution operators lead to generalized Liouville equations that produce norm-preserving dynamics of $\brho$. The mathematical form of these equations are similar to the quantum von-Neumann-Liouville's equation.

We have also shown that the generator $\stability_-$ produces an evolution in the form of a  ``classical Bloch sphere'' of phase space volume that is time invariant. 
In addition, we showed that the framework for this $\jacobian_-$ based evolution is particularly useful for chaotic systems in which tangent vectors evolving under the linearized dynamics collapse on to the most expanding direction at a give phase point.
Because $\jacobian_-$ generates a unitary (orthogonal) evolution, the inner product of tangent vectors remains preserved. 
This continuous orthogonal evolution avoids the discrete reorthonormalization steps of the standard Gram-Schmidt/QR procedure, while cleanly separating rotational and stretching effects at the generator level.

	This modified dynamics provides a continuous orthogonal evolution that avoids the need for periodic GS reorthonormalization when maintaining an orthonormal tangent frame.
Any orthogonal set of basis vectors remains so throughout the evolution, by construction, and the instantaneous Lyapunov exponents are computed using the symmetric part of the stability matrix $\stability_+$.
As examples, we considered several set of basis states and numerical computation of their finite-time exponents for Hamiltonian and non-Hamiltonian systems. 

We showed that there is a relationship between the antisymmetric evolution and the continuous QR/Gram–Schmidt procedure: QR dynamics combines the physical rotation generated by $\stability_-$ with an additional constraint-induced rotation arising from $\stability_+$, whereas the present splitting isolates the intrinsic rotational component. This separation of stretching and rotation at the generator level provides a geometric framework for analyzing tangent-space dynamics that is independent of Hamiltonian structure. We anticipate that this perspective may prove useful in studying invariant subspace geometry, angle-based diagnostics, and the structure of chaotic attractors in higher-dimensional systems.

\begin{acknowledgments}
This material is based upon work supported by the National Science Foundation under Grant No. 2124510.
\end{acknowledgments}

\vfill
\newpage

\section*{Appendix}

\appendix
\renewcommand{\theequation}{\thesubsection\arabic{equation}}
\setcounter{equation}{0}
\renewcommand{\thesubsection}{\Alph{subsection}}

\subsection{Properties of the evolution matrix \texorpdfstring{$\widetilde{\bm{M}}$}{M}}
\label{SI:evol-mat}
\subsubsection*{Norm-preserving}

Along a phase space trajectory, a unit perturbation vector $\ket{\delta \yu}$ evolves in the tangent space according to:
\begin{align}
d_t\ket{\delta\yu} = \stability\ket{\delta\yu}  - \langle\stability_+\rangle\ket{\delta\yu},
\end{align}
where $\stability = \stability_+ + \stability_-$ and $\langle\stability_+\rangle = \bra{\delta\yu} \stability_+\ket{\delta\yu} $. The solution of this equation of motion is given by:
\begin{align}
\ket{\delta\yu(t)} = \widetilde{\jacobian}\ket{\delta\yu(t_0)}.
\end{align}
Here, $\widetilde{\jacobian}$ is a the norm-preserving operator given by
\begin{align}
\widetilde{\jacobian}(t,t_0) =\frac{\mathcal{T}e^{\int_{t_0}^t\,\stability_-dt'}\mathcal{T}e^{\int_{t_0}^t\,\stability_+dt'}}{\mathcal{T}\langle e^{\int_{t_0}^t2\stability_+dt'}\rangle_{t_0}^{1/2}}.
\end{align}
where $\mathcal{T}$ stands for time-ordered integration indicating the continuum limit of successive  multiplications.

To see that $\widetilde\jacobian$ recovers the equation of motion for $\ket{\delta\yu}$, we differentiate it with respect to $t$:
\begin{align}
d_t\widetilde{\jacobian} = \stability_-\widetilde{\jacobian}+\stability_+\widetilde{\jacobian} - \langle\stability_+\rangle\widetilde{\jacobian}.
\end{align}
This operator $\widetilde{\jacobian}$ is norm-preserving as it preserves the unit norm for $\ket{\delta \yu(t_0)}$ as it evolves with time.

\subsubsection*{Normal but non-orthogonal}

The operator $\widetilde{\jacobian}$ is normal  $\widetilde{\jacobian}^\top\widetilde{\jacobian}$= $\widetilde{\jacobian}\widetilde{\jacobian}^\top$. To see this, we write these matrix product below:
\begin{align}
	\widetilde{\jacobian}^\top \widetilde{\jacobian}&= \frac{\mathcal{T'}e^{\int_{t_0}^t\,\stability_+dt'}\mathcal{T}e^{\int_{t_0}^t\,\stability_+dt'}}{\mathcal{T}\langle e^{\int_{t_0}^t2\stability_+dt'}\rangle_{t_0}} \nonumber\\
	&=  \frac{\mathcal{T}e^{\int_{t_0}^t\,2\stability_+dt'}}{\mathcal{T}\langle e^{\int_{t_0}^t2\stability_+dt'}\rangle} = \widetilde{\jacobian} \widetilde{\jacobian}^\top.
\end{align}

The exponential terms with $\stability_-$ cancel out as time-ordering is taken into account. 
Next, the expectation value 
\begin{align}
\langle\widetilde{\jacobian}^\top\widetilde{\jacobian}\rangle= \langle\widetilde{\jacobian}\widetilde{\jacobian}^\top\rangle = 1,
\end{align}
which indicates the norm-preserving property of $\widetilde{\jacobian}$.

However, $\widetilde{\jacobian}$ is non-orthogonal in general:  
\begin{align}
\widetilde{\jacobian}^\top\widetilde{\jacobian}= \widetilde{\jacobian}\widetilde{\jacobian}^\top \neq \mathbbm{1}. 
 \end{align}
 
\subsection{Volume preserving operator \texorpdfstring{$\widetilde{\bm{M}}'$}{M}}\label{SI_vol}
The operator $\widetilde{\jacobian}$ (or $\widetilde{\jacobian_+}$) is norm-preserving and specific to a pure state. However, for maximally mixed state which is given by a equally weighted sum of pure states formed using the complete set of $n$ basis vectors, we would need a set of $n$ operators $\widetilde{\jacobian}_i$. 

To derive the Generalized Liouville's equation in the main text, we construct a volume preserving operator $\widetilde{\jacobian}'$ given by:
\begin{align}
	\widetilde{\jacobian}' = \left(\prod_{i=1}^n\widetilde{\jacobian}_i\right)^{1/n},
\end{align}
where
\begin{align}
\widetilde{\jacobian}_i(t,t_0) =\frac{ \mathcal{T}e^{\int_{t_0}^t\,\stability_-dt'}\mathcal{T}e^{\int_{t_0}^t\,\stability_+dt'}}{\langle\mathcal{T}e^{2\int_{t_0}^t\stability_+dt'}\rangle_{i,t_0}^{1/2}}.
\end{align} 
Moreover, we see that for $\delta t \to 0$:
\begin{align}
	\mathcal{T}\langle e^{\int_{t_0}^t2\stability_+dt'}\rangle_{t_0} \approx \mathcal{T} e^{\int_{t_0}^t2\langle\stability_+\rangle dt'}, \nonumber
\end{align}
which leads to:
\begin{align}
\widetilde{\jacobian}(t,t_0) =\frac{\mathcal{T}e^{\int_{t_0}^t\,\stability_-dt'}\mathcal{T}e^{\int_{t_0}^t\,\stability_+dt'}}{\mathcal{T} e^{\int_{t_0}^t\langle\stability_+\rangle_{i} dt'}}. \nonumber
\end{align}

Thus, the determinant of volume preserving $\widetilde{\jacobian}'$ becomes:
\begin{align}
	|\widetilde{\jacobian}'|^{1/n} =  \mathcal{T} e^{-\int_{t_0}^t\sum_{i=1}^n\langle\stability_+\rangle_i dt'}\mathcal{T}e^{\int_{t_0}^t\,\Tr\stability_+dt'}, \nonumber
\end{align} 
where we have used the relation $|e^{\boldsymbol{X}}| = e^{\Tr\boldsymbol{X}}$ for a square matrix $\boldsymbol{X}$ and the identity $\Tr \stability_- = 0$ as $\stability_-$ is anti-symmetric.

But $\sum_{i=1}^n\langle\stability_+\rangle_i$ is the sum of instantaneous Lyapunov exponents which is the phase space volume contraction rate $\Lambda = \Tr \stability_+$. So, the terms $-\sum_{i=1}^n\langle\stability_+\rangle_i$ and $\Tr \stability_+$ in argument of the resulting exponential precisely cancel out and we find:
\begin{align}
	|\widetilde{\jacobian}'| = 1.
\end{align} 
The same treatment  for the symmetric operator $\widetilde{\jacobian}_+' = \sum_{i,j=1}^n[\delta_{ij}]\widetilde{\jacobian}_{+i}$ leads to $|\widetilde{\jacobian}_+'|=1$. Therefore, both of these volume preserving operators belong to $SL(n)$.

\subsection{Derivation of the Liouville equation for the maximally mixed state}
\label{SI:LE-derivation}
Here we show that the determinant of the maximally mixed density matrix
$\brho_{\mathrm{max}}$ is preserved under the normalized evolution,
\begin{equation}
d_t \ln |\brho_{\mathrm{max}}|
= \Tr\!\left(d_t \ln \brho_{\mathrm{max}}\right)
= 0 .
\end{equation}

We begin with the definition of the maximally mixed state,
\begin{equation}
\brho_{\mathrm{max}}
= \frac{1}{n}\sum_{i=1}^{n} \brho_i ,
\end{equation}
where $\{\ket{\delta\bm{\phi}_i(t)}\}_{i=1}^n$ is a complete set of tangent-space
basis vectors spanning an $n$-dimensional phase-space volume element
$d\mathcal{V}(t)$ at time $t$, and
\begin{equation}
\brho_i = \dyad{\delta\bm{\phi}_i}{\delta\bm{\phi}_i}
\end{equation}
are the corresponding pure-state matrices.

The expectation value of the stability matrix $\stability$ with respect to
$\brho_{\mathrm{max}}$ is
\begin{align}
\langle \stability \rangle_{\brho_{\mathrm{max}}}
&= \Tr(\stability \brho_{\mathrm{max}})
= \frac{1}{n}\sum_{i=1}^n \Tr(\stability \brho_i) \nonumber = \frac{1}{n}\Lambda ,
\end{align}
where $ \Lambda = \sum_{i=1}^n \Tr(\stability \brho_i)$
is the instantaneous phase-space volume contraction rate (the sum of local
Lyapunov exponents).

We now compute the time derivative of the logarithm of the determinant of
$\brho_{\mathrm{max}}$.
Using $\Tr(d_t \ln \brho) = \Tr(\brho^{-1} d_t \brho)$ for an invertible matrix,
we obtain:
\begin{align}
d_t \ln |\brho_{\mathrm{max}}|
&= \Tr\!\left(\brho_{\mathrm{max}}^{-1} d_t \brho_{\mathrm{max}}\right).
\end{align}
The equation of motion for $\brho_{\mathrm{max}}$ under the normalized dynamics
reads:
\begin{equation}
d_t \brho_{\mathrm{max}}
= \{\bar{\stability}, \brho_{\mathrm{max}}\}
+ [\stability_-, \brho_{\mathrm{max}}],
\end{equation}
where $\bar{\stability} = \stability - \langle \stability \rangle_{\brho_{\mathrm{max}}}\mathbbm{I}$.
Substituting this expression yields
\begin{align}
d_t \ln |\brho_{\mathrm{max}}|
&= \Tr\!\left(
\brho_{\mathrm{max}}^{-1} \{\bar{\stability}, \brho_{\mathrm{max}}\}
\right)
+ \Tr\!\left(
\brho_{\mathrm{max}}^{-1} [\stability_-, \brho_{\mathrm{max}}]
\right).
\end{align}
The commutator term vanishes identically under the trace,
\begin{equation}
\Tr\!\left(
\brho_{\mathrm{max}}^{-1} [\stability_-, \brho_{\mathrm{max}}]
\right) = 0 ,
\end{equation}
while the anticommutator term gives
\begin{align}
\Tr\!\left(
\brho_{\mathrm{max}}^{-1} \{\bar{\stability}, \brho_{\mathrm{max}}\}
\right)
&= 2\Tr(\bar{\stability}) \nonumber \\
&= 2\left[\Tr(\stability) - n\langle \stability \rangle_{\brho_{\mathrm{max}}}\right] \nonumber \\
&= 2(\Lambda - \Lambda) = 0 .
\end{align}
Combining the two contributions, we conclude that
\begin{equation}
d_t \ln |\brho_{\mathrm{max}}| = 0 ,
\end{equation}
demonstrating that the determinant of the maximally mixed density matrix, and
hence the associated phase-space volume element, is preserved under the
normalized evolution.

\subsection{Cancellation of the anti-commutator term in the \texorpdfstring{$\stability_+$}{A+} eigenbasis}
\label{SI:A-basis-EOM}
We clarify here the conditions under which the equation of motion simplifies when expressed in a basis adapted to the symmetric stability matrix $\stability_+$.

Let $\brho_i = \dyad{\delta\bm{\phi}_i}{\delta\bm{\phi}_i}$ denote a pure-state projector onto an eigenvector of $\stability_+$, such that
\begin{equation}
\stability_+ \brho_i = \sigma_i \brho_i,
\qquad
\langle \stability_+ \rangle
= \Tr(\stability_+ \brho_i)
= \sigma_i,
\end{equation}
where $\sigma_i$ is the eigenvalue of $\stability_+$ associated with the eigenvector $\ket{\delta\bm{\phi}_i}$.

Using the definition
\begin{equation}
\bar{\stability}_+ = \stability_+ - \langle \stability_+ \rangle \mathbb{1},
\end{equation}
the anti-commutator appearing in the equation of motion becomes
\begin{align}
\{\bar{\stability}_+,\brho_i\}
&= \{\stability_+,\brho_i\}
   - 2\langle \stability_+ \rangle \brho_i \nonumber\\
&= (\sigma_i \brho_i + \sigma_i \brho_i)
   - 2\sigma_i \brho_i \nonumber\\
&= \mathbb{0}.
\end{align}
Thus, when the density matrix $\brho_i$ is evaluated in the eigenbasis of $\stability_+$, the anti-commutator term vanishes identically. This cancellation is basis dependent and does not hold for an arbitrary choice of basis. The commutator term, by contrast, does not generally vanish and governs the tangent-space evolution.

\subsection{Continuous QR/Gram-Schmidt procedure and antisymmetric evolution}
\label{SI:comparison}
We clarify the relation between antisymmetric evolution and the
continuous limit of the QR method.

\subsection*{Pure antisymmetric evolution}

Let $\boldsymbol{A}=\boldsymbol{A}_+ + \boldsymbol{A}_-$ denote the symmetric and
antisymmetric decomposition of the stability matrix,
with $\boldsymbol{A}_+^\top=\boldsymbol{A}_+$ and
$\boldsymbol{A}_-^\top=-\boldsymbol{A}_-$.
If an orthonormal basis $\boldsymbol{Q}$ evolves according to
\begin{equation}
    \dot{\boldsymbol{Q}} = \boldsymbol{A}_- \boldsymbol{Q},
\end{equation}
then
\begin{equation}
    \frac{d}{dt}(\boldsymbol{Q}^\top \boldsymbol{Q})
    =
    \boldsymbol{Q}^\top \boldsymbol{A}_- \boldsymbol{Q}
    +
    (\boldsymbol{Q}^\top \boldsymbol{A}_- \boldsymbol{Q})^\top
    =
    0.
\end{equation}
Thus orthonormality is preserved exactly.
This evolution represents pure rotation and discards all stretching
information carried by $\boldsymbol{A}_+$.

\subsection*{Continuous QR dynamics}

In the QR method, one evolves a set of vectors $\boldsymbol{V}$:
\begin{equation}
    \dot{\boldsymbol{V}} = \boldsymbol{A}\boldsymbol{V},
    \qquad
    \boldsymbol{V}=\boldsymbol{Q}\boldsymbol{R},
\end{equation}
where $\boldsymbol{Q}^\top\boldsymbol{Q}=\mathbb{1}$ and $\boldsymbol{R}$ is upper
triangular.
Differentiating $\boldsymbol{V}=\boldsymbol{Q}\boldsymbol{R}$ gives
\begin{equation}
    \boldsymbol{A}\boldsymbol{Q}\boldsymbol{R}
    =
    \dot{\boldsymbol{Q}}\boldsymbol{R}
    +
    \boldsymbol{Q}\dot{\boldsymbol{R}}.
\end{equation}
Multiplying on the left by $\boldsymbol{Q}^\top$ and on the right by
$\boldsymbol{R}^{-1}$,
\begin{equation}
\boldsymbol{Q}^\top\dot{\boldsymbol{Q}}
=
\boldsymbol{Q}^\top \boldsymbol{A}\boldsymbol{Q}
-
\dot{\boldsymbol{R}}\boldsymbol{R}^{-1}.
\label{eq:QRcompact}
\end{equation}
Since $\boldsymbol{Q}^\top\boldsymbol{Q}=\mathbb{1}$,
\begin{equation}
    \boldsymbol{Q}^\top\dot{\boldsymbol{Q}}
    =
    -(\boldsymbol{Q}^\top\dot{\boldsymbol{Q}})^\top,
\end{equation}
so define the skew-symmetric generator
\begin{equation}
    \boldsymbol{K}:=\boldsymbol{Q}^\top\dot{\boldsymbol{Q}},
    \qquad
    \dot{\boldsymbol{Q}}=\boldsymbol{Q}\boldsymbol{K}.
\end{equation}
Because $\dot{\boldsymbol{R}}\boldsymbol{R}^{-1}$ is upper triangular,
Eq.~\eqref{eq:QRcompact} uniquely decomposes
$\boldsymbol{Q}^\top\boldsymbol{A}\boldsymbol{Q}$
into skew-symmetric and upper-triangular parts.
Hence
\begin{equation}
    \boldsymbol{K}
    =
    \operatorname{skew}\!\left(\boldsymbol{Q}^\top\boldsymbol{A}\boldsymbol{Q}\right),
\end{equation}
where
\begin{equation}
    \operatorname{skew}(\boldsymbol{X})
    :=
    \frac12(\boldsymbol{X}-\boldsymbol{X}^\top).
\end{equation}

\subsection*{Decomposition of the generator}

Using $\boldsymbol{A}=\boldsymbol{A}_+ + \boldsymbol{A}_-$,
\begin{align}
\boldsymbol{K}
&=
\operatorname{skew}\!\left(
\boldsymbol{Q}^\top \boldsymbol{A}_+ \boldsymbol{Q}
+
\boldsymbol{Q}^\top \boldsymbol{A}_- \boldsymbol{Q}
\right) \\
&=
\operatorname{skew}\!\left(\boldsymbol{Q}^\top \boldsymbol{A}_+ \boldsymbol{Q}\right)
+
\boldsymbol{Q}^\top \boldsymbol{A}_- \boldsymbol{Q}.
\end{align}
Therefore,
\begin{equation}
    \dot{\boldsymbol{Q}}
    =
    \boldsymbol{A}_-\boldsymbol{Q}
    +
    \boldsymbol{Q}\,
    \operatorname{skew}\!\left(\boldsymbol{Q}^\top \boldsymbol{A}_+ \boldsymbol{Q}\right).
\end{equation}
The first term is the physical rotation generated by the antisymmetric
part of the stability matrix.
The second term is a constraint-induced rotation originating from the
symmetric part, required to maintain orthonormality while stretching
occurs.
Thus, antisymmetric evolution and continuous QR coincide only when
\begin{equation}
\operatorname{skew}(\boldsymbol{Q}^\top \boldsymbol{A}_+ \boldsymbol{Q})=0,
\end{equation}
i.e., in special cases where the symmetric part produces no rotational component in the moving frame.

\bibliography{references}
\end{document}